\documentclass[12pt,recno]{JHEP3}

\usepackage{graphicx,amsfonts,amssymb,stmaryrd,amsmath,mondef,longtable}
\usepackage[all]{xy}

\advance\footskip by -0.8cm

\title{Defects and D-Brane Monodromies}  
\author{Ilka Brunner$^{1}$,
Hans Jockers$^{2}$ and 
Daniel Roggenkamp$^{3}$ \\  
\\
$^{1}$Institut f{\"u}r Theoretische Physik, ETH Z\"urich \\
CH--8093 Z{\"u}rich, Switzerland\\ 
E-mail: \email{brunner@itp.phys.ethz.ch}\\
\\
$^{2}$Stanford Institute for Theoretical Physics\\
Stanford CA 94305-4060, USA\\
E-mail: \email{jockers@stanford.edu}\\
\\
$^{3}$Department of Physics and Astronomy, Rutgers University\\
Piscataway, NJ 08855-0849, USA\\
E-mail: \email{roggenka@physics.rutgers.edu}
}
\abstract{In this paper D-brane monodromies are studied from a world-sheet point of view.
More precisely, defect lines are used to describe the parallel transport of D-branes along deformations
of the underlying bulk conformal field theories. This method is used to derive B-brane monodromies in K\"ahler moduli spaces of non-linear sigma models on projective hypersurfaces. The corresponding defects are constructed at Landau-Ginzburg points in these moduli spaces where matrix factorisation techniques can be used. Transporting them to the large volume phase by means of the gauged linear sigma model we find that their action on B-branes at large volume can be described by certain Fourier-Mukai transformations which are known from target space geometric considerations to represent the corresponding monodromies.}
\preprint{RUNHETC-2008-12\\
SU-ITP-08/11}
\begin{document}
\section{Introduction}
As is well known, D-branes in a given closed string background
form a category, with morphisms given by open strings stretching from one D-brane to another.
An important tool in the study of such D-brane categories
are functors between them.
Examples of such functors are T-duality, mirror symmetry and the monodromies
generated by transporting D-branes  
around singular points in moduli spaces of Calabi-Yau compactifications.

From a world-sheet point of view, D-branes are described by boundary conditions
in the conformal field theory associated to the closed string vacuum under consideration, 
and natural operations on them arise from defects. Defect lines are one-dimensional junctions separating 
possibly different conformal field theories\footnote{We do not distinguish defects between non-isomorphic CFTs, by calling them ``interfaces'' as is \eg done in \cite{Bachas:2007td}.} on the same world-sheet (see \eg \cite{Petkova:2000ip,Bachas:2001vj,Frohlich:2004ef,Bachas:2004sy,Frohlich:2006ch,Quella:2006de,Alekseev:2007in,Fuchs:2007tx,Runkel:2007wd} for more details on defects in conformal field theories).
Pushing such a defect line onto world-sheet boundaries, a new boundary condition is created out of the one originally imposed there.
For general defects this fusion process is singular (see \eg \cite{Bachas:2007td} for an example in the free boson CFT), 
but if defects and boundary conditions preserve
the same $N=2$ superconformal symmetry, it can be regularised and effectively described in the topologically 
twisted theory \cite{Brunner:2007qu, Brunner:2007ur}. In this way, supersymmetry preserving defects give rise to functors between D-brane categories. 

The functors we will consider in this article are monodromy transformations obtained by transporting 
B-type D-branes around singularities in K\"ahler moduli space of $N=2$ supersymmetric non-linear $\sigma$-models. This in particular includes the monodromies around Landau-Ginzburg and conifold points. 

By means of target space geometric methods those functors have been derived in \cite{Horja,Aspinwall:2001dz,Distler:2002ym,Aspinwall:2002nw} following a proposal by Kontsevich \cite{Rutgerslecture}.
Using the large volume realisation of B-branes, they can be represented in terms
Fourier-Mukai transformations on the bounded derived category of coherent sheaves on the target space manifold, $D^b(X)$.

The approach we will take in this article is more direct. We propose to use world-sheet techniques to analyse the
behaviour of B-type D-branes, \ie B-type boundary conditions, under deformations of the underlying bulk conformal field theories along closed loops in its K\"ahler moduli spaces. This will be studied using a novel method put forward in \cite{Brunner:2007ur}. Namely, on general grounds one expects that the effect of K\"ahler perturbations of bulk conformal field theories on B-branes can be described by the fusion with associated B-type supersymmetry preserving defects between the deformed and the undeformed theories.

Thus, our task 
is to find the specific B-type defects associated to deformations along given loops around singularities in K\"ahler moduli spaces. 
As base points for these loops
we choose the Landau-Ginzburg or Gepner points which exist for the models we will consider.
At these points use can be made of an elegant description of B-branes and B-type defects in terms of matrix factorisations \cite{Kapustin:2002bi,Brunner:2003dc,Khovanov:2004bc,Brunner:2007qu}. 

To compare the resulting defects to the Fourier-Mukai transformations obtained using target space geometric methods,
we need to transport them from the Landau-Ginzburg point in K\"ahler moduli space to the large volume limit. 
The functor realising this transport on the level of B-type D-branes has been constructed in \cite{Orlov} and has recently been given a physical understanding in \cite{Herbst:2008jq} (see also \cite{Aspinwall:2006ib}). In the latter work, B-type D-branes in gauged linear $\sigma$-models have been defined and studied. This provides a uniform description of B-branes on those parts of K\"ahler moduli space on which the non-linear $\sigma$-models admit a gauged linear $\sigma$-model realisation.
In particular, the techniques developed in \cite{Herbst:2008jq} can be used to transport B-type branes inside these parts 
of K\"ahler moduli space, and they immediately generalise to the treatment of B-type supersymmetry preserving defects. 

We therefore restrict our considerations to non-linear $\sigma$-models on projective hypersurfaces, which can be realised as gauged linear $\sigma$-models. B-type defects defined at the Landau-Ginzburg point can then be lifted to the gauged linear $\sigma$-model and transported to the large volume limit. As a general result we find that any B-type defect
transported from the Landau-Ginzburg point to large volume in this way gives rise to a Fourier-Mukai transformation
on the large volume realisation $D^b(X)$ of the B-brane category. 

Such a relation is not completely unexpected. After all, the ``folding
trick'' allows to interpret defects between two conformal field
theories as boundary conditions in the tensor product of the first of
these theories with the (parity dual) of the other one
\cite{Oshikawa:1996dj,Bachas:2001vj}, and also Fourier-Mukai
transformations between two large volume B-brane cateogries are
specified by B-branes on the product of the respective target space
manifolds.

The Fourier-Mukai transformations that arise  by  transporting 
the monodromy defects we construct at the Landau-Ginzburg point to large volume
indeed agree with the ones obtained by more geometric methods in \cite{Horja,Aspinwall:2001dz,Aspinwall:2002nw}.

In fact, lifting defects to the gauged linear $\sigma$-models and transporting them in the respective K\"ahler moduli spaces
provides an independent way to derive the defects describing monodromies generated by loops in the those parts of K\"ahler moduli space which are covered by a GLSM description.
Namely, by lifting the identity defect to the gauged linear $\sigma$-model, transporting it around a singularity in the K\"ahler moduli space and then pushing it down to the non-linear $\sigma$-model again, one obtains the respective monodromy defect. This will be used as a confirmation of our more general construction of monodromy defects at the Landau-Ginzburg point. 

The plan of this paper is as follows:
In the remaining part of this introduction, we review the monodromies in
K\"ahler moduli space of one-parameter models in Section \ref{sec:reviewI} and give some details about
the fusion of supersymmetric defects in Section \ref{defectsection}. Following \cite{Brunner:2007ur}, we will motivate
that the effect K\"ahler deformations of bulk theories have on B-branes can be described by the fusion with B-type defects. 
In Section \ref{defectmfsection}
we review the matrix factorisation formalism used to describe B-type D-branes and defects at Landau-Ginzburg points.
The monodromy defects will be constructed in this framework. 
In Section \ref{glsm} we outline how B-type D-branes and defects can be transported on K\"ahler moduli space by lifting them to the gauged linear $\sigma$-model. We only give a brief sketch of the constructions we need and illustrate them in the example of tensor product B-branes. For detailed explanations we refer the reader to \cite{Herbst:2008jq}. 
Using these techniques, we argue in 
Section \ref{defecttransport} that upon transport from the Landau-Ginzburg point, B-type 
defects generally give rise to Fourier-Mukai transformations at large volume.
Section \ref{LGmonsection} is devoted to the construction and the analysis of the defects corresponding to the monodromy around Landau-Ginzburg points. At the Landau-Ginzburg models themselves, they are obtained as those defects realising the quantum symmetries which these models exhibit.
Transporting them to large volume, we determine the respective Fourier-Mukai kernels. Furthermore, transporting the identity defect around closed loops in K\"ahler moduli space we also 
obtain the defects and Fourier-Mukai transformations for monodromies around those singularities which can be encircled in the K\"ahler moduli space of the corresponding gauged linear $\sigma$-model. 
In Section \ref{conifoldsection} we construct general conifold monodromies in Landau-Ginzburg models and transport them to large volume. 
Finally, in Section \ref{discussionsection} we conclude with some remarks about directions of future research.
\subsection{D-brane Monodromies in K\"ahler moduli space}\label{sec:reviewI}
The aim of this paper is to 
study the dependence of B-type D-branes on K\"ahler parameters using world-sheet methods. 
In particular, we are interested
in monodromies of B-type D-branes around singular points in the K\"ahler moduli space.
We will focus our considerations on non-linear $\sigma$-models with Calabi-Yau target spaces, which can be realised as 
hypersurfaces in projective space such as the cubic torus in $\IP^2$ or 
the quintic hypersurface in $\IP^4$.

Such non-linear $\sigma$-models admit a realisation as gauged linear $\sigma$-models with ${\rm U}(1)$ gauge group \cite{Witten:1993yc}. The corresponding K\"ahler moduli space\footnote{In general the K\"ahler moduli spaces of the gauged linear $\sigma$-models only capture two-dimensional submanifolds of the non-linear $\sigma$-model K\"ahler moduli spaces. In case of projective hypersurfaces of dimension three and higher the Lefschetz hyperplane theorem guarantees that the two K\"ahler moduli spaces coincide.} 
is given by a two sphere which however has three singular points. 
The large volume point is the limiting point where the volume of the target space goes to infinity. Similarly, at the conifold point
the (quantum) volume of the target space goes to zero. Finally, the Landau-Ginzburg or Gepner point is an orbifold point in the K\"ahler moduli space. At this point certain dualities of the theories under consideration become self-dualities and enhance the symmetry group of the theory.

We would like to study what happens to B-type D-branes when one moves around these singular points in K\"ahler moduli space.
As alluded to above, from a world-sheet point of view 
this means that one has to understand how boundary conditions evolve under deformations of the bulk conformal field theories.  

For a general CFT however, boundary conditions are very difficult to construct. Even worse, in the situations we are interested in, the CFTs themselves are only known explicitly at special points in the K\"ahler moduli spaces and conformal perturbation theory becomes 
intractable at higher orders in the perturbation parameters. 

Fortunately, there are relatively nice descriptions of B-type branes at special points in the K\"ahler moduli space. 
At large volume for instance B-type D-branes have a geometric realisation in terms of bounded complexes of coherent sheaves on the target space 
\cite{Douglas:2000gi,Lazaroiu:2001jm,Aspinwall:2001pu}, and as mentioned above,  using target space geometric arguments one can determine their monodromies around the singular points in K\"ahler moduli space.

Namely, it has been been shown how the conifold monodromy manifests itself as a result
of stability considerations in \cite{Aspinwall:2001dz}. The basic physical idea underlying these arguments 
is that at a conifold point the quantum world-volume of a particular D-brane collapses to zero. The D-brane
becomes massless \cite{Strominger:1995cz} and its central charge goes to zero at this point.
Therefore, this D-branes can be created at the conifold point at no cost in energy.
The quantum world-volume of other D-branes and therefore their masses remain finite.
If such a massive probe brane is transported around the
conifold point, the open string states between the probe brane 
and the D-brane which is massless at the conifold point
become tachyonic at some point, triggering a tachyon condensation of these D-branes
to an energetically preferable bound state.
Thus, taking the probe brane around the conifold point, one ends up with a bound state of it with
the D-brane which becomes massless at the conifold point \cite{Aspinwall:2001dz,Aspinwall:2004jr,Jockers:2006sm}.

Moreover, it is not difficult to see that the large volume monodromy can be understood in terms of a shift in the B-field
$B \to B+\xi$, where $\xi$ is the generator\footnote{If $\dim H^{1,1}(X)>1$ we will only consider the monodromy around the large volume limit in the slice of the moduli space covered by the GLSM construction. Then $\xi$ is the generator of
$H^{1,1}(X)$ inherited from the one-dimensional cohomology group $H^2(\PP^{N-1},\mathbb{R})$ of the embedding projective space.} of $H^{1,1}(X)$.
Since the B-field enters the relevant charge formulas for large volume B-branes 
only in the combination $B+F$ with the field strength $F$ of the ${\rm U}(1)$-connection on the D-brane,
the action  of the monodromy is given by tensoring the Chan-Paton bundle of the B-brane
with a line bundle of first Chern class $c_1=[\xi]$ \footnote{Note that the appearance of $F+B$ in the charge formulas a priori justifies the action of the monodromy on the level of K-theory. However, by studying B-type D-branes in the context of open topological string field
theory \cite{Lazaroiu:2001jm,Diaconescu:2001ze}, this action carries over to the derived category.} 

The monodromy around the Landau-Ginzburg point is not so easy to derive directly by means of geometric arguments. However, it
can be obtained by composing conifold and large volume monodromies. A more direct derivation of this monodromy appeared in \cite{Mayr:2000as}, combining linear $\sigma$-model methods for the open string with the fact that this monodromy is given by the quantum symmetry at the Landau-Ginzburg point.

To derive these monodromies directly using world-sheet techniques, 
one has to deform the bulk conformal field theory along loops in K\"ahler moduli space and 
analyse how B-type boundary conditions behave under these deformations. The straight forward way to approach this 
problem is to use conformal perturbation theory, which however becomes intractable at higher orders. 
Therefore we will employ a method put forward in \cite{Brunner:2007ur} to analyse the effect of bulk perturbations on boundary conditions. Namely, we will use the fact that K\"ahler perturbations can be described by B-type defects between the unperturbed and the perturbed theory. This will be explained in more detail in Section \ref{defectsection} below.
\subsection{Bulk Perturbations and Defects} \label{defectsection}
In this section we will briefly outline some aspects of defects and their fusion. Furthermore, we will
motivate why the effect of bulk perturbations of conformal field theories on boundary conditions can be described by the fusion with defects between the unperturbed and the perturbed theory.

\FIGURE{
\includegraphics[height=6.5cm]{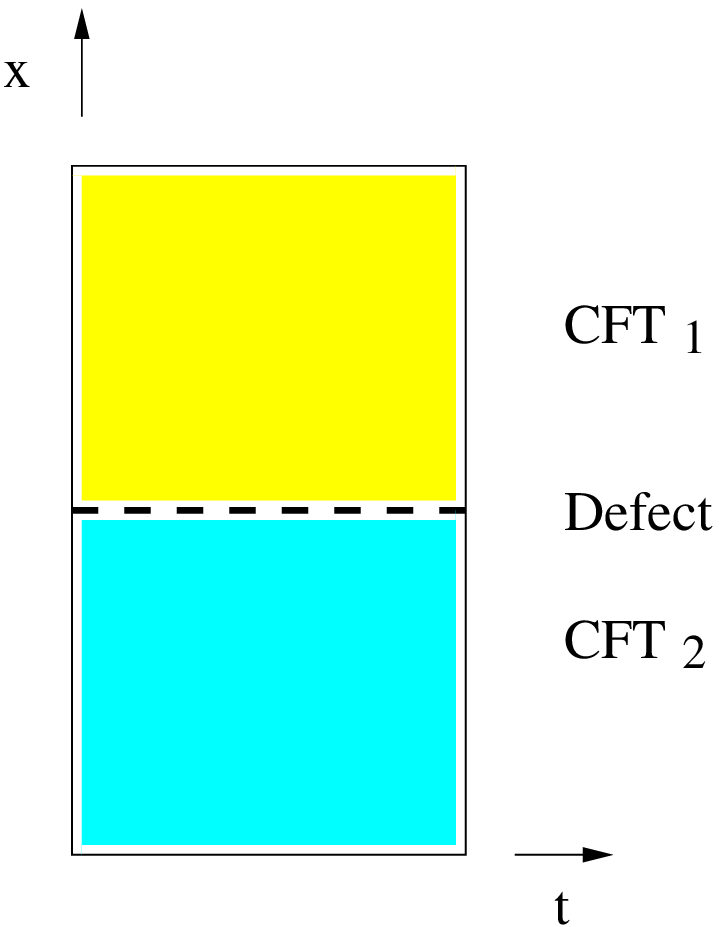}
\caption{\small Two two-dimensional CFTs are separated by a
one-dimensional defect.}\label{fig:interface}}
As mentioned above, defects are one-dim\-en\-sional junctions separating two possibly different conformal field
theories on the same world-sheet (\cf Figure \ref{fig:interface}).

Two such defects between conformal field theories can be brought on top of each other to form a new defect
as illustrated in Figure \ref{fig:fusion}. 
Such a ``fusion'' process involves taking
the limit in which the distance between the defects shrinks to zero. On
the level of the full conformal field theory this process is highly singular in general
and needs to be regularised (see \cite{Bachas:2007td} for an explicit
calculation in the example of the free boson CFT). In case the underlying conformal field theories
are $N=2$ superconformal, and the defects preserve the same $N=2$ superconformal symmetry, 
this fusion can be regularised. Indeed, it can be described on the level of the topologically twisted theory,
where correlators do not depend on positions on the world-sheet, in particular not on those of defects. 
Hence, in the topologically twisted theory, correlators do not exhibit singularities when defects approach each other, and fusion is well defined. 
\FIGURE{
\includegraphics[height=5cm]{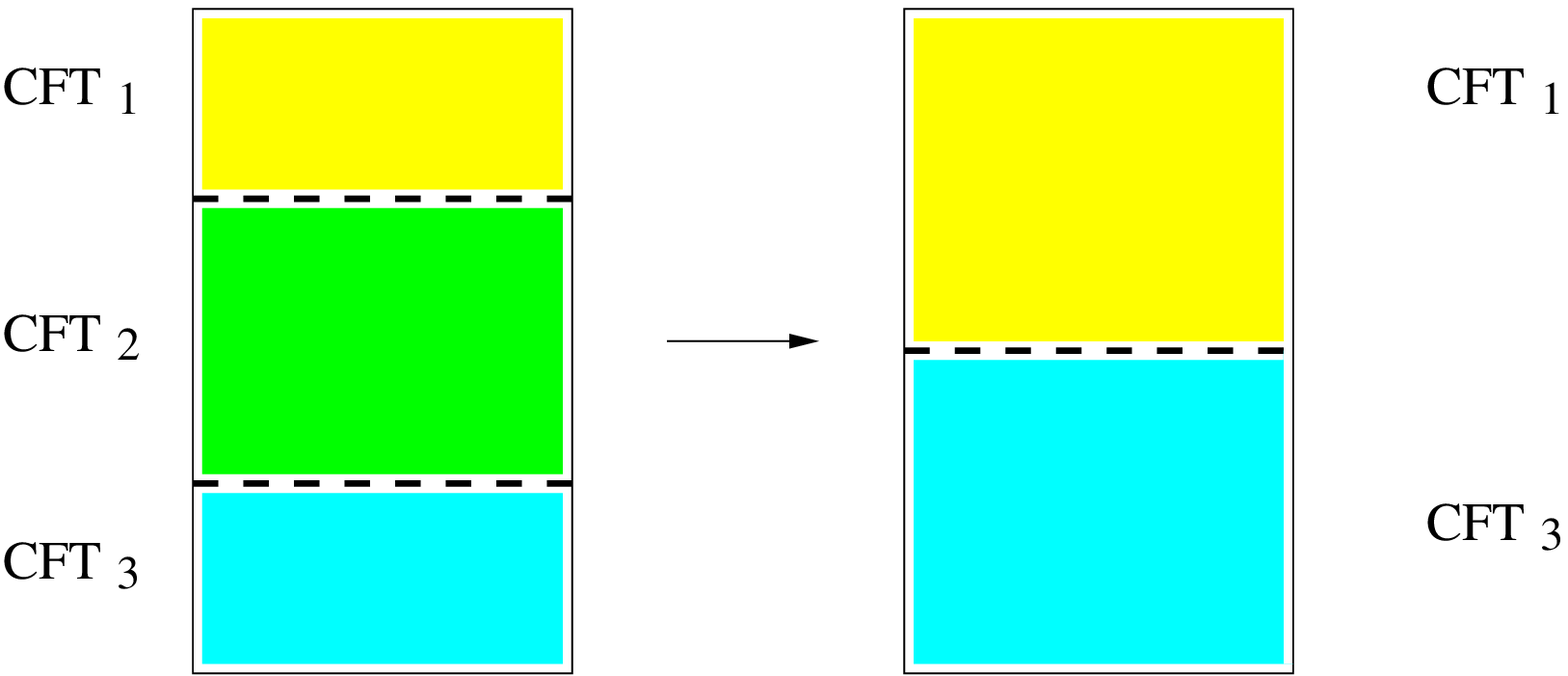}
\caption{\small In the limit where the two defects coincide, a new
defect connecting theory 1 and 3 emerges.}\label{fig:fusion}}

In the same way, defects can be brought on top of world-sheet boundaries to transform the original boundary conditions
imposed there into new ones (\cf Figure \ref{fig:bdryfuse}). In general, also the fusion of a defect and a boundary condition is singular, but as in the case of fusion of two defects, it can be regularised if defect and boundary conditions preserve the same $N=2$ superconformal symmetry. (In fact, boundary conditions in a conformal field theory can be regarded as defects between this CFT and the trivial one, so that boundary condition are in fact a special type of defects.)
\FIGURE{
\includegraphics[height=5cm]{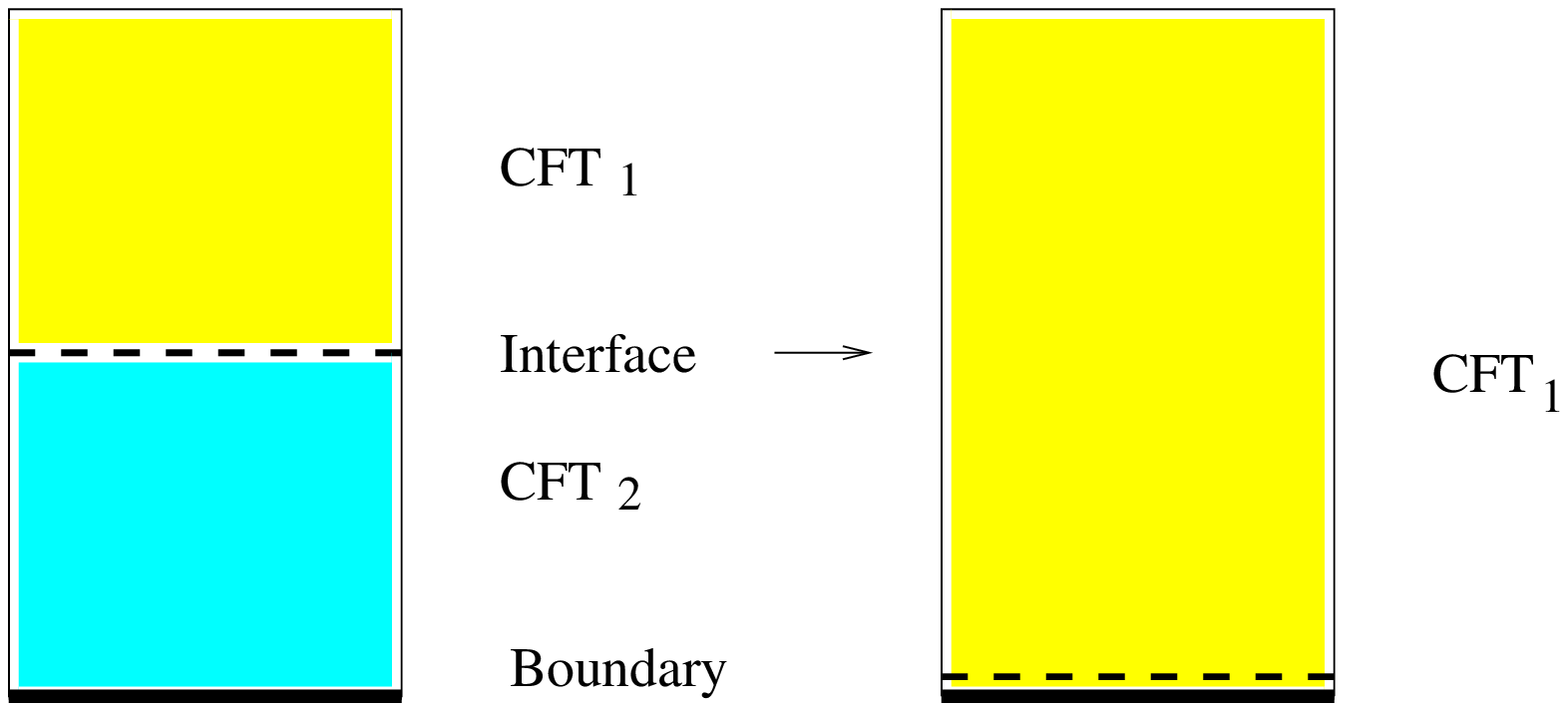}
\caption{\small In the limit where the defect coincides with the boundary, a new
boundary condition of theory 1 emerges.}\label{fig:bdryfuse}}

In this way, fusion endows the set of all defects preserving a specific $N=2$ superconformal symmetry with a multiplicative structure, and these defects act on the category of D-branes preserving the same $N=2$ algebra by means of functors between the respective D-brane categories. For the case of Landau-Ginzburg theories, fusion of B-type defects and their action on boundary conditions has been investigated in \cite{Brunner:2007qu,Brunner:2007ur}.

Analogously to D-branes, also defects exhibit fields which are defined on them. These ``defect-changing fields'' can be used to perturb defects in the same way as ``boundary condition-changing fields'' can be used to perturb boundary conditions. In particular, also defects can undergo tachyon condensation and bound state formation. Not surprisingly, such bound states will figure prominently in the construction of defects describing monodromies around conifold points.

Defects have a rich structure and are very interesting objects, and they also turn out to be very useful. For instance,
it has been realised in \cite{Brunner:2007ur} that they can be employed 
to study the behaviour of D-branes under (marginal or relevant) 
perturbations of the underlying bulk conformal field theory\footnote{Boundary
flows have been realised by means of defects in \cite{Bachas:2004sy,Alekseev:2007in}.}. 

The basic reasoning behind this is to 
restrict the perturbation of a conformal field theory on a world-sheet with boundary a fixed distance $\epsilon$ away from the latter.
Performing the renormalisation group flow associated to this restricted perturbation, one ends up with the following situation in the IR (\cf Figure \ref{fig:flowdefect}). On the part of the surface the perturbation was restricted to, the theory has flown from the UV to the IR of the particular bulk perturbation considered, whereas the theory in the $\epsilon$-strip around the boundary from which the perturbation was kept away stays at the UV. These two theories are separated by a defect line situated in $\epsilon$-distance from the boundary. In this way, every bulk perturbation gives rise to a unique ``flow defect''.
\FIGURE{
\includegraphics[height=4.6cm]{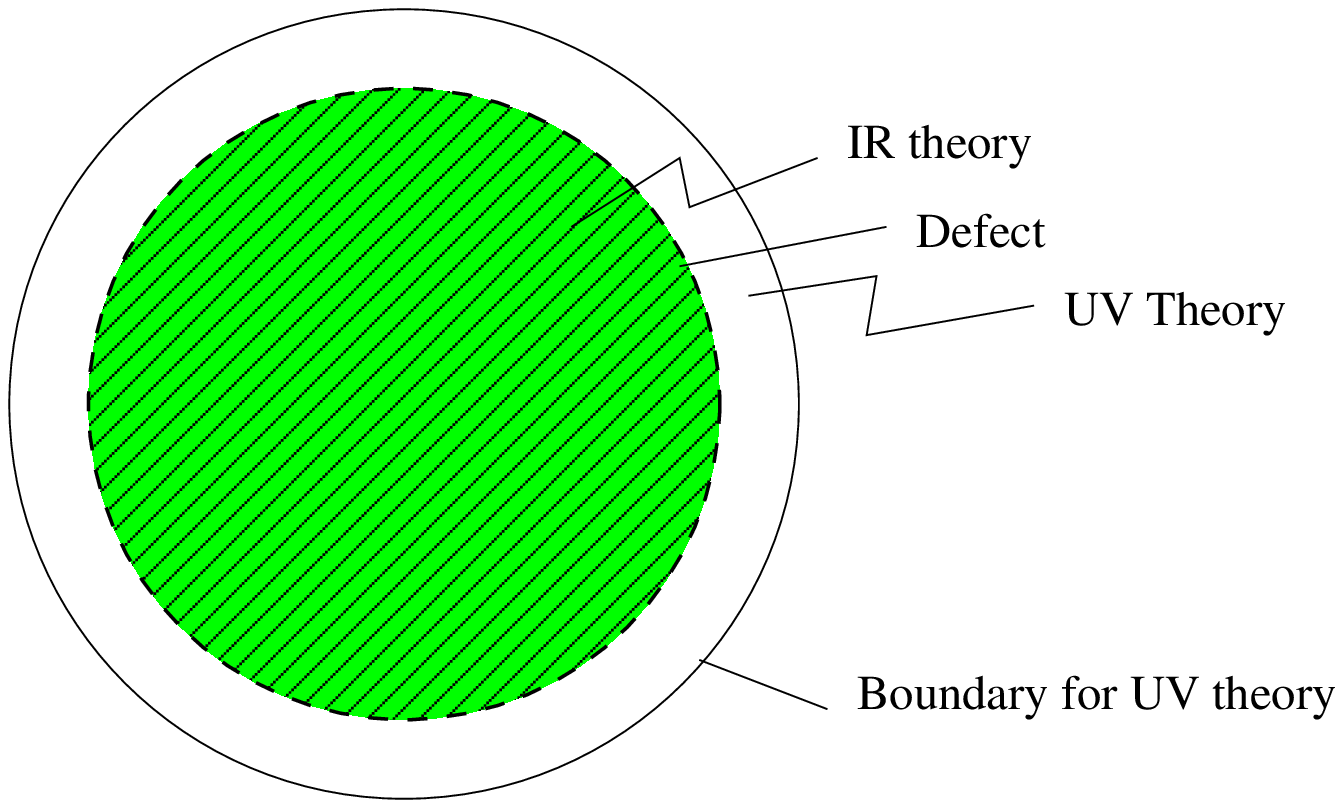}
\caption{\small The perturbation is restricted to the shaded domain. UV and IR
are separated by a defect. Fusing this defects with the UV boundary,
one obtains the IR boundary condition.}\label{fig:flowdefect}}

Moving this defect towards the boundary, \ie sending $\epsilon$ to zero, spreads the perturbation to the entire surface. 
As alluded to above this process is singular in general. That is not surprising, because the perturbing bulk fields can have singularities when they approach the boundary. For that reason the unrestricted bulk perturbation on the surface with boundary needs to be regularised and counterterms on the boundary have to be added. In general such an unrestricted bulk perturbation will therefore also trigger a simultaneous boundary perturbation.
In the same way, the process of taking the defect towards the boundary needs to be regularised. The regularised fusion of the defect with the boundary condition then describes to what boundary condition in the IR theory a given boundary condition in the UV theory flows under the bulk perturbation, \ie it encodes the behaviour of boundary conditions under bulk flows. 

The corresponding regularisation can be very tricky however. Fortunately, we are interested in a special case, namely bulk perturbations of $N=(2,2)$ superconformal field theories, which preserve an $N=2$ supersymmetry even in the presence of world-sheet boundaries. For these perturbations, one can topologically twist the corresponding theories all along the flow. The flow defects also preserve the $N=2$ supersymmetry in this case, and its fusion with boundary conditions can be effectively described in the twisted theory, where no singularities
arise when the defect is moved towards the boundary. In this way, it is possible to avoid dealing with regularisation issues for these particular perturbations. 

The condition that the perturbation preserves supersymmetry on a surface with boundary however is very constraining.
There are two classes of $N=(2,2)$ supersymmetry preserving bulk perturbations, those generated by elements of the 
$(c,c)$- and $(a,a)$-rings (chiral multiplets) which correspond to ``complex structure-type'' perturbations, and those generated by elements of the $(a,c)$- and $(c,a)$-rings (twisted-chiral multiplets), which correspond to ``K\"ahler-type'' perturbations. 
Concretely, the
bulk perturbation generated by a twisted chiral field takes the form (notations are taken from \cite{Hori:2000ic})
\beq
\Delta S= \int d\bar{\theta}^- d \theta^+ \Psi \vert_{\bar{\theta}^+=\theta^-=0}
+ c.c. \, ,
\eeq
where $\Psi$ is the perturbing twisted chiral multiplet. 
On world-sheets without boundary, the perturbation is invariant under the
full $N=(2,2)$-supersymmetry variation
\beq\label{susydelta}
\delta = \epsilon_+ Q_- - \epsilon_- Q_+ - \bar{\epsilon}_+ \bar{Q}_-
+ \bar{\epsilon}_- \bar{Q}_+ \ .
\eeq
On world-sheets with boundaries one can impose A- or B-type boundary
conditions and preserve one half of the supersymmetry.
An A-type supersymmetry variation (setting $\epsilon:=\epsilon_+ =\bar{\epsilon}_-$ in \eq{susydelta}) of a
twisted chiral perturbation 
yields a boundary term that in general cannot be cancelled by the supersymmetry variation of a boundary term\footnote{In fact, this calculation is mirror to the one performed in \cite{Kapustin:2002bi,Brunner:2003dc,Hori:2000ic} in the case of B-type
boundary conditions for the superpotential part of the action. There, additional degrees of freedom at the boundary helped to cancel the undesired term. Here, we can only use the fields that are already part of the theory.}.
Hence, in the presence of a boundary on which an A-type boundary condition is imposed, the 
bulk perturbation generated by a twisted chiral field breaks supersymmetry.

On the other hand, the boundary term coming from
a B-type supersymmetry variation (\ie $\epsilon:=\epsilon_+=-\epsilon_-$, 
$\bar{\epsilon}:=\bar{\epsilon}_+=-\bar{\epsilon}_-$)
can always be cancelled by adding a simple additional boundary
term involving only bulk fields
\beq
{\cal B} \sim \int_{\partial \Sigma} (\psi - \bar\psi) ,
\eeq
where $\psi$ denotes the upper component of the twisted chiral superfield $\Psi$.
Thus, twisted chiral bulk perturbations preserve supersymmetry in the presence of boundaries with B-type boundary conditions\footnote{Note however that the perturbations can rotate the gluing conditions imposed on the spectral flow operator. On the level of CFT this can
be seen using the methods of \cite{Fredenhagen:2006dn}. From a space time point
of view, this implies that the target space supersymmetry is rotated and
possibly broken. These effects were fundamental for the discussion of $\Pi$-stability.}.

Hence, it can be concluded that K\"ahler-type perturbations preserve supersymmetry in the presence
of boundary conditions of B-type but not in general of those of A-type. The flow defects associated to these perturbations
preserve the same B-type supersymmetry.
Similarly, complex structure-type perturbations preserve supersymmetry in the presence of boundary conditions of 
A-type but not of those of B-type, and the corresponding flow defects preserve A-type supersymmetry. 

In both these cases regularisation issues in the fusion of flow defects with boundary conditions can be avoided by considering the fusion in the topologically twisted theory, the B-twisted one in the first case, and the A-twisted one in the second.

In this article we are interested in the first of these two cases. Namely, from a world-sheet perspective, monodromies of B-branes in the K\"ahler moduli space of non-linear $\sigma$-models are obtained by deforming the non-linear $\sigma$-model
in the presence of B-type boundary conditions along closed loops in K\"ahler moduli space. By the reasoning above the effect of this deformation on B-branes can be described by the fusion with a B-type flow-defect. 
\section{B-type Defects in Landau-Ginzburg Models and Matrix Factorisations}\label{defectmfsection}
In Landau-Ginzburg models both, B-type supersymmetric D-branes as well as B-type supersymmetry preserving defects have an elegant description
in terms of matrix factorisations \cite{Kapustin:2002bi,Brunner:2003dc,Khovanov:2004bc,Brunner:2007qu}. A matrix factorisation $P$ of a polynomial
$W\in\CC[x_1,\ldots,x_N]$ is given by a pair $(P_1,P_0)$ of free $\CC[x_1,\ldots,x_N]$ modules together with homomorphisms $p_s:P_s\rightarrow P_{(s+1)\,{\rm mod}\, 2}$ between them 
which compose to $W$ times the identity map, \ie $p_1p_0=W\id_{P_0}$ and $p_0p_1=W\id_{P_1}$. In the following we will often represent matrix factorisations by
\beq
P: P_1\overset{p_1}{\underset{p_0}{\rightleftarrows}}P_0\,.
\eeq
Sometimes it is useful to regard them as two-periodic twisted\footnote{The differential squares to $W$.} complexes. Indeed, such matrix factorisations form a category, with morphisms $\HH^*(P,Q)$ between two matrix factorisations $P$ and $Q$ given by the cohomology of the $Hom$-complex of the two twisted complexes associated to $P$ and $Q$. The latter is a two-periodic untwisted complex.  

There are always matrix factorisations with modules $P_s=\CC[x_1,\ldots,x_N]$ where one of the maps $p_r=1$ and the other one
$p_{(r+1)\,{\rm mod}\,2}=W$. They are trivial in that they only have zero-morphisms with any other (including themselves) matrix factorisation, and two matrix factorisations which differ by the addition of such a trivial matrix factorisation are equivalent. 

As was shown in \cite{Kapustin:2002bi,Brunner:2003dc}, B-type supersymmetric D-branes in Landau-Ginzburg models with chiral superfields $x_1,\ldots,x_N$ and 
superpotential $W\in\CC[x_1,\ldots,x_N]$ can be represented by matrix factorisations of $W$, where open strings between two such D-branes are described by morphisms
between the respective matrix factorisations. 

In the same way, it has been argued in \cite{Brunner:2007qu} that B-type supersymmetry preserving defects between two Landau-Ginzburg models, one with chiral fields
$x_1,\ldots,x_N$ and superpotential $W_1\in\CC[x_1,\ldots,x_N]$ and one with chiral superfields $y_1,\ldots,y_M$ and superpotential $W_2\in\CC[y_1,\ldots,y_M]$
can be represented by matrix factorisations of $W_1-W_2$ over the polynomial ring $\CC[x_1,\ldots,x_N,y_1,\ldots,y_N]$.

As mentioned in Section~\ref{defectsection}, one interesting property of $N=2$-supersymmetric defects is that they can be fused with other such defects, or with D-branes (boundary conditions) preserving the same supersymmetry. Namely, two such defects can be brought on top of each other to produce a new defect, or a defect can be moved onto a world sheet boundary to change the boundary condition imposed there.
This fusion has a very simple realisation in terms of the matrix factorisation description. For instance, let $x_i$, $y_i$, $z_i$ be the chiral superfields of three Landau-Ginzburg models with superpotentials $W_1\in\CC[x_i]$, $W_2\in\CC[y_i]$ and $W_3\in\CC[z_i]$ respectively, which are separated by two 
defects represented by matrix factorisations $P^1$ of $W_1-W_2$ and $P^2$ of $W_2-W_3$. Fusing the two defects gives rise to a new defect separating the Landau-Ginzburg model with chiral fields $x_i$ and superpotential $W_1$ from the one with chiral fields $z_i$ and superpotential $W_3$. This fused defect is given by the matrix factorisation
\beq
P^1*P^2=\left(P^1\otimes P^2\right)_{\CC[x_i,z_i]}^{\rm red}\,.
\eeq
Here, the tensor product of two matrix factorisations is defined by taking the tensor product of the associated twisted complexes, which again is a two-periodic complex which is twisted by the sum of the twists of the tensor factors. More concretely, the tensor product $P\otimes Q$ of matrix factorisations $P$ and $Q$ of $W$ and $W\p$ respectively can be written as
\beq\label{eq:MFtensor}
P\otimes Q: P_1\otimes Q_0\oplus P_0\otimes Q_1\overset{r_1}{\underset{r_0}{\rightleftarrows}}P_0\otimes Q_0\oplus P_1\otimes Q_1
\eeq
with 
\beq
r_1=\left(\begin{array}{cc} p_1\otimes\id&\id\otimes q_1\\-\id\otimes q_0&p_0\otimes\id\end{array}\right)\,,\quad
r_0=\left(\begin{array}{cc} p_0\otimes\id&-\id\otimes q_1\\ \id\otimes q_0&p_1\otimes\id\end{array}\right)
\eeq
which is a matrix factorisation of $W+W\p$.

In the situation above, $P^1$ is a matrix factorisation of $W_1-W_2$ and $P^2$ one of $W_2-W_3$.
Hence, $P^1\otimes P^2$ is a matrix factorisation of $W_1-W_3\in\CC[x_i,z_i]$, but it is still a matrix factorisation over $\CC[x_i,y_i,z_i]$. That means that the modules $(P^1\otimes P^2)_s$ are free $\CC[x_i,y_i,z_i]$-modules and also the maps 
$r_s$ between them depend on the $y_i$. 
The notation $\left(P^1\otimes P^2\right)_{\CC[x_i,z_i]}$ means that this matrix factorisation has to be regarded as one over $\CC[x_i,z_i]$ only. As such, it is of infinite rank, because the modules $(P^1\otimes P^2)_s$ regarded as modules over $\CC[x_i,z_i]$ 
are free modules of infinite rank. For instance, $\CC[x_i,y_i,z_i]$ can be decomposed as 
\beq
\CC[x_i,y_i,z_i]=\bigoplus_{(l_1,\ldots,l_N)\in\NN_0^N}y_1^{l_1}\ldots y_N^{l_N}\CC[x_i,z_i]
\eeq
into free $\CC[x_i,z_i]$-modules.
Physically speaking, the chiral fields $y_i$ of the theory squeezed in between the two defects are promoted to new defect degrees of freedom in the limit where the two defects coincide. However, most of them are trivial. Namely, if both $P^1$ and $P^2$ are of finite rank, 
the matrix factorisation $\left(P^1\otimes P^2\right)_{\CC[x_i,z_i]}$ can be reduced to finite rank by splitting off infinitely many trivial matrix factorisations. It is the result of this reduction $\left(P^1\otimes P^2\right)_{\CC[x_i,z_i]}^{\rm red}$ which describes the fused defect. More details about this can be found in \cite{Brunner:2007qu}.

In the same way, fusion of B-type defects and B-type D-branes in Landau-Ginzburg models can be formulated in the matrix factorisation framework. The fusion of a B-type defect separating a Landau-Ginzburg model with chiral fields $x_i$ and superpotential $W_1\in\CC[x_i]$ from one with chiral fields $y_i$ and superpotential $W_2\in\CC[y_i]$ and a B-type D-brane (boundary condition) in the second of these Landau-Ginzburg models can be represented by the matrix factorisation
\beq
P*Q=\left(P\otimes Q\right)_{\CC[x_i]}^{\rm red}\,,
\eeq
where $P$ is the matrix factorisation of $W_1-W_2$ associated to the defect and $Q$ the matrix factorisation of $W_2$ associated to the D-brane.

This formalism can be easily generalised to the context of Landau-Ginzburg orbifolds.
If $\Gamma$ is an orbifold group in a Landau-Ginzburg model with chiral fields $x_1,\ldots,x_N$ and superpotential $W\in\CC[x_1,\ldots,x_N]$, then 
it has an action $\rho:\Gamma\rightarrow{\rm End}(\CC[x_1,\ldots,x_N])$ on $\CC[x_1,\ldots,x_N]$ which is compatible with multiplication, \ie
\beq\label{orbact}
\rho(\gamma)(ab)=\rho(\gamma)(a)\rho(\gamma)(b)\quad{\rm for}\;a,b\in\CC[x_1,\ldots,x_N],\,\gamma\in\Gamma\,,
\eeq
and which leaves $W$ invariant. The Landau-Ginzburg orbifolds which will appear in this article are of the following type. 
$W\in\CC[x_1,\ldots,x_N]$ is a homogeneous polynomial of degree $d$, and the 
orbifold group $\Gamma=\ZZ_d$ acts on monomials of degree $l$ by phase multiplications:
\beq
\gamma(n)(x_1^{l_1}\ldots x_N^{l_N})=e^{\frac{2\pi i}{d}n\sum_il_i} x_1^{l_1}\ldots x_N^{l_N}\,\quad n\in\ZZ_d\,.
\eeq
B-type D-branes in Landau-Ginzburg orbifolds can then be represented by $\Gamma$-equivariant matrix factorisations of $W$ \cite{Ashok:2004zb,Hori:2004ja}. 
These are matrix factorisations
\beq
P: P_1\overset{p_1}{\underset{p_0}{\rightleftarrows}}P_0
\eeq
together with representations $\rho_s$ of $\Gamma$ on $P_s$, which are compatible with the module structure
\beq
\rho_s(\gamma)(ar)=\rho(\gamma)(a)\rho_s(\gamma)(r)\quad{\rm for}\;a\in\CC[x_1,\ldots,x_N],\,\gamma\in\Gamma,\,r\in P_s
\eeq
and commute with the maps $p_s$
\beq\label{commrel}
\rho_{(s+1)\,{\rm mod}\,2}(\gamma)p_s=p_s\rho_s(\gamma)\quad{\rm for}\;\gamma\in\Gamma,\, s=0,1\,.
\eeq
Note that because of the compatibility with the multiplication, the $\rho_i$ are fixed by the respective action on the degree-zero subspaces $\CC^k\subset\CC[x_i]^k\cong P_s$. 
It is this action which we will later specify in the definition of equivariant matrix factorisations. In the case of $\ZZ_d$- or ${\rm U}(1)$-actions we will often indicate the respective charge $q$ of $1\in\CC[x_i]$ by writing $\CC[x_i][q]$. 

Given two such equivariant matrix factorisations $P$ and $Q$ the above conditions ensure that the $\Gamma$-action pushes through to the cohomology of the $Hom$-complex, so that one can define the space of morphism in the category of equivariant matrix factorisations to be $\HH_{\rm orb}^*(P,Q)=(\HH^*(P,Q))^\Gamma$, the $\Gamma$-invariant part of the morphism spaces in the underlying unorbifolded category. These are the spaces of open strings between the respective D-branes in the orbifold Landau-Ginzburg model.

Similarly, B-type defects separating a Landau-Ginzburg orbifold with chiral fields $x_i$, superpotential $W_1\in\CC[x_i]$ and orbifold group $\Gamma_1$ and one
with chiral fields $y_i$, superpotential $W_2\in\CC[y_i]$ and orbifold group $\Gamma_2$ can be represented by $\Gamma_1\times\Gamma_2$-equivariant matrix factorisations of $W_1-W_2$ \cite{Brunner:2007ur}. The conditions on the $\Gamma$-representations above imply that the matrix factorisation obtained 
by fusing two equivariant matrix factorisations $P$ and $Q$ in the unorbifolded category is again equivariant with respect to the product of all orbifold groups present. 
In particular, one can take the part of this matrix factorisation which is invariant under the orbifold group $\Gamma_{\rm squeezed}$ of the model which is squeezed in between the two defects, or the defect and the boundary. This represents the fusion in the orbifold theory
\beq\label{eq:OrbFusion} 
P*_{\rm orb}Q=\left(P*Q\right)^{\Gamma_{\rm squeezed}}\,.
\eeq
Since most of the discussion in the following will be concerned with orbifolds of Landau-Ginzburg models, 
we will often omit to write the subscript `orb' in case confusion is unlikely.

As alluded to above, in this article we are dealing with homogeneous superpotentials $W$ of degree $d$. The corresponding Landau-Ginzburg models exhibit a ${\rm U}(1)_R$-symmetry which acts on the chiral fields as
\beq
\rho^R(\varphi)(x_1^{l_1}\ldots x_N^{l_N})=e^{2\pi i\varphi\sum_i \frac{2l_i}{d}}x_1^{l_1}\ldots x_N^{l_N}\,\quad \varphi\in\RR\,,
\eeq
\ie the $x_i$ have ${\rm U}(1)_R$-charge $\frac{2}{d}$, such that the superpotential has charge $2$.
The existence of this additional symmetry guarantees that the theory flows to a conformal field theory in the IR. 
Similarly, also B-type boundary conditions and defects of the Landau-Ginzburg model flow to conformal boundary conditions and defects in the IR if they preserve this ${\rm U}(1)_R$-symmetry. This means that they are given by ${\rm U}(1)_R$-equivariant matrix factorisations. 
Since $W$ is not invariant under ${\rm U}(1)_R$, but has $R$-charge $2$, one has to slightly modify the definition of equivariant matrix factorisations in this case. More precisely the commutation relation \eq{commrel} cannot hold in this case, but it has to be replaced by
\beq
\rho^R_{(s+1)\,{\rm mod}\,2}(e^{2\pi i\varphi}) p_s =e^{2\pi i\varphi}p_s\rho^R_s(e^{2\pi i\varphi})\,,
\eeq
reflecting the fact that the maps $p_s$ have $R$-charge $1$. 
Note that there is a relation between the representation $\rho^R$ of ${\rm U}(1)_R$ and the representation $\rho$ of the orbifold group $\Gamma=\ZZ_d$. First of all the representation of ${\rm U}(1)_R$ and $\Gamma=\ZZ_d$ commute with each other. Secondly, the combination $\rho^R(\frac{1}{2})\rho(-1)$ leaves all chiral fields $x_i$ invariant. Thus, for irreducible matrix factorisations $P$, $\rho^R_s(\frac{1}{2})\rho_s(-1)\sim\id_{P_s}$. Furthermore, $\rho^R_s(\frac{1}{2})\rho_s(-1)$ anti-commute with the maps $p_s$, and it can be chosen to be
\beq
\rho^R_s\left({\s \frac{1}{2}}\right)\rho_s(-1)=(-1)^s\id_{P_s}\,.
\eeq
Diagonalising the actions of $\rho^R_s$ and $\rho_s$ simultaneously, one arrives at the following relation between the ${\rm U}(1)_R$-charges $r$ and $\Gamma=\ZZ_d$-charges $q$ on the modules $P_s$
\beq
r-\frac{2q}{d}\in 2\ZZ+s\;\;\;{\rm on}\, P_s\,.
\eeq
\section{Gauged Linear Sigma Models and Matrix Factorisations}\label{glsm}
To compare the monodromies we obtain in terms of defects at the Landau-Ginzburg points with the monodromy transformations
at large volume derived by more geometric methods \cite{Aspinwall:2002nw,Aspinwall:2004jr}, we need to transport B-type D-branes between these two points in K\"ahler moduli space. The corresponding functors between the categories of B-branes in Landau-Ginzburg models and those in non-linear $\sigma$-models has been constructed in \cite{Orlov}. A physical understanding of these functors has been developed in \cite{Herbst:2008jq}, using a gauged linear $\sigma$-model description of B-type D-branes, see \cite{Aspinwall:2006ib,Mayr:2000as,Govindarajan:2000vi,tomasiello:2000ym} for earlier work.

Indeed, the models we are considering have a realisation in terms of gauged linear $\sigma$-models with gauge group ${\rm U}(1)$,
chiral superfields $(\pf,x_1,\ldots,x_N)$ of ${\rm U}(1)$-charge $(-d,1,\ldots,1)$ and $R$-charge $(2,0,\ldots,0)$, and superpotential
$\pf W(x_1,\ldots,x_N)$, where $W$ is a polynomial of degree $d$ in the fields $x_i$ (see \cite{Witten:1993yc}). We are mostly interested in the Calabi-Yau case, in which $d=N$. These models depend on a complex parameter $t=r+i\theta$, a combination of the Fayet-Iliopoulus parameter $r$ and the $\theta$-angle, which parametrises their K\"ahler moduli spaces $(r,\theta)\in(\RR\times\RR/2\pi\ZZ)\backslash\{(d\log d,[\pi d])\}$. The large volume limit point is 
situated at $t=\infty$, where the GLSM reduces to a $\sigma$-model on the projective hypersurface $X\subset\PP^{N-1}$ defined by $W=0$. At $t=-\infty$ on the other hand the field $\pf$ receives a non-vanishing vacuum expectation value, which breaks the gauge group down to $\ZZ_d$, and the GLSM reduces to a Landau-Ginzburg orbifold with chiral superfields $x_i$, superpotential $W$ and orbifold group $\Gamma=\ZZ_d$. The conifold point corresponds to $(r=d\log d,\theta=[\pi d])$. 

The GLSMs provide a uniform description of all the models in K\"ahler moduli space. A description of D-branes in these models therefore allows to transport B-type D-branes between arbitrary points in K\"ahler moduli space. This has been worked out in \cite{Herbst:2008jq}, and will be used here to compare monodromies at Landau-Ginzburg and large volume points. 
In this section we will briefly describe the techniques we need. For more details we refer the reader to \cite{Herbst:2008jq}.
\subsection{B-type D-branes in GLSMs}
Because of the monodromies B-type D-branes exhibit on the K\"ahler moduli spaces of gauged linear $\sigma$-models, a uniform description of these D-branes can only be expected on a cover of these moduli spaces. In the cases we are considering, this cover is given by $\RR\times\RR\backslash(\{d\log d\}\times 2\pi\ZZ+\pi d)$. On it, 
B-type D-branes can be represented by ${\rm U}(1)\times{\rm U}(1)_R$-equivariant matrix factorisations of $\pf W(x_1,\ldots,x_N)$. However, on the phase boundary $r=d\log d$ not all of these matrix factorisations lead to physically well defined boundary conditions. This gives rise to the ``grade restriction rule'' of \cite{Herbst:2008jq}. 
Namely, on a component $N_n=\{d\log d\}\times(\pi d-2\pi (n+1)+(0,2\pi))$ of the phase boundary, only matrix factorisations
give rise to well defined boundary conditions, whose ${\rm U}(1)$-charges $q$ satisfy
\beq
1-d+n\leq q\leq n\,.
\eeq
We denote the respective ``window'' of integral ${\rm U}(1)$-charges compatible with this phase boundary component $\NNN_n=\{1-d+n,\ldots,n\}$.

Our aim is to transport B-branes from the Landau-Ginzburg to the large volume point of such a GLSM, where they can be represented by matrix factorisations of $W$ and complexes of coherent sheaves on $X$ respectively. To this end, we will discuss in the following how B-branes behave under the reduction of the GLSM to these limiting models. We will however only present the recipe here. More detailed explanations can be found in \cite{Herbst:2008jq}.
\subsection{Reduction to Landau-Ginzburg models} \label{sec:GLSMtoLG}
In the Landau-Ginzburg phase of the GLSM the field $\pf$ receives a non-vanishing vacuum expectation value \cite{Witten:1993yc}, which can be gauged to $1$. This breaks the gauge group from ${\rm U}(1)$ down to $\ZZ_d$. The fields $x_i$ remain massless and the model reduces to the Landau-Ginzburg orbifold with superpotential $W$ and orbifold group $\Gamma=\ZZ_d$. 

On the level of B-type D-branes, a ${\rm U}(1)\times{\rm U}(1)_R$-equivariant matrix factorisation $\widehat P$
representing a B-type brane in the GLSM reduces to a $\ZZ_d\times{\rm U}(1)_R$-equivariant matrix factorisation of $W$, representing a B-type brane in the Landau-Ginzburg orbifold in the following way.
One sets all entries $\pf$ in the maps $\widehat p_s$ to $1$ and replaces $\widehat P_s$ by $\widehat P_s/\pf\widehat P_s$. This reduces the ${\rm U}(1)$-equivariance to a $\ZZ_d$-equivariance, where the representations $\widehat\rho_s$ are replaced by the induced representations of $\ZZ_d\subset{\rm U}(1)$. 
Furthermore, the ${\rm U}(1)_R$-representations $\rho^R_s$ of the reduced matrix factorisations arise as the combination $\rho_s^R=\widehat \rho^R_s\widehat\rho_s^2$ of the ${\rm U}(1)_R$ and ${\rm U}(1)$-representations of the GLSM matrix factorisations.

Since we would like to transport B-type branes from the Landau-Ginzburg to the large volume point, we are interested in the reverse of this reduction. Namely, given a B-brane in the Landau-Ginzburg orbifold, \ie a $\ZZ_d\times{\rm U}(1)_R$-equivariant matrix factorisation $P$ of $W$, we would like to lift it to a ${\rm U}(1)\times{\rm U}(1)_R$-equivariant matrix factorisation $\widehat P$ of $\pf W$, which can be transported through a specified component $N_n$ of the phase boundary, and which by the above procedure reduces to $P$ at the Landau-Ginzburg point. Indeed, in the Calabi-Yau case, for every matrix factorisation $P$ of $W$ and every component $N_n$ of the phase boundary, there is exactly one such lift $\widehat P$ \cite{Herbst:2008jq}. The lifts can be constructed by introducing factors $\pf$ into the maps $p_s$ in such a way that the representations $\rho_s$ of $\ZZ_d$ lift to representations $\widehat\rho_s$ of ${\rm U}(1)$ and that the corresponding charges lie in the charge window $\NNN_n$. We refrain from giving more details about the general construction here, but will later on explicitly discuss the lifts we will need. In particular, the case of tensor product B-branes is discussed in Section \ref{tpbranes} below.
\subsection{Reduction to Large Volume}\label{sec:GLSMtoLV}
At large volume, one recovers the complex of coherent sheaves representing the large volume B-brane out of the GLSM matrix factorisation $\widehat P$ in the following way. Consider the module
${\mathcal P}=\widehat P_0\oplus\widehat P_1/W(\widehat P_0\oplus \widehat P_1)$ and regard it as 
$R=\CC[x_1,\ldots,x_N]/(W)$-module. Because of the presence of the $\pf$, it is a free $R$-module of infinite rank, but since $\pf$ has ${\rm U}(1)_R$-charge $2$, every submodule ${\mathcal P}^r$ of fixed ${\rm U}(1)_R$-charge $r\in\ZZ$ is a free $R$-module of finite rank. Moreover, there is an $r_{\rm min}\in\ZZ$, such that for all $r<r_{\rm min}$ ${\mathcal P}^r=0$, and, having ${\rm U}(1)_R$-degree $1$, the maps $\widehat{p}_s$ define 
maps ${\mathcal P}^r\rightarrow{\mathcal P}^{r+1}$. In this way, one obtains from $\widehat P$ 
a ${\rm U}(1)$-equivariant complex 
\beq
{\mathcal P}:{\mathcal P}^{r_{\rm min}}\rightarrow{\mathcal P}^{r_{\rm min}+1}\rightarrow
{\mathcal P}^{r_{\rm min}+1}\rightarrow\ldots
\eeq
of $R$-modules, in which the position in the complex is determined by the ${\rm U}(1)_R$-charge.
This complex is bounded to the left, but unbounded to the right. The desired complex $\widetilde{\mathcal P}$ 
of coherent sheaves 
on the projective hypersurface $X={\rm Proj}(R)$ in $\PP^{N-1}$ is obtained from ${\mathcal P}$ by sheafification. In particular, free $R$-modules $R[q]$ with ${\rm U}(1)$-representation specified by $q\in\ZZ$
which appear in ${\mathcal P}$ give rise to sheaves $\OO_X(-q)$ on $X$ in $\widetilde{\mathcal P}$. A priori this is a complex of coherent sheaves on $X$ which is unbounded to the right, but in fact it is quasi-isomorphic to a bounded complex \cite{Herbst:2008jq}. In this way, one obtains the large volume interpretation of the GLSM B-brane $\widehat P$. In the next subsection we will present a simple example of this construction.
\subsection{Example: Tensor Product Branes}\label{tpbranes}
As an example let us discuss how to transport tensor product B-branes from the Landau-Ginzburg to the large volume phase.
These B-branes exist for any Landau-Ginzburg model with homogeneous superpotential. So let $W\in S:=\CC[x_1,\ldots,x_N]$ be a homogeneous polynomial of degree $d$. 
Then $W$ can be written as
\beq
W(x_i)=\sum_j x_j A_j(x_i)
\eeq
for some homogeneous polynomials $A_j$ of degree $d-1$. The $A_j$ are not unique in general, but the equivalence class of the tensor product $P=\bigotimes_j P(j)$ of rank-one matrix factorisations
\beq
P(j): P(j)_1\cong S\overset{p(j)_1=x_j}{\underset{p(j)_0=A_j}{\rightleftarrows}}S\cong P(j)_0
\eeq
of $x_jA_j$ is independent of the choice. Being a tensor product matrix factorisation, 
\beq
P: P_1\overset{p_1}{\underset{p_0}{\rightleftarrows}}P_0
\eeq
has the following form
\beqn\label{tpmaps}
&&P_s=\mathop{\bigoplus_{(s_i)\in\ZZ_2^N}}_{s-\sum_i s_i\,{\rm even}}\bigotimes_{i=1}^NP(i)_{s_i}\,,\\
&&p_s=
\mathop{\sum_{(s_i)\in\ZZ_2^N}}_{s-\sum_i s_i\,{\rm even}}\sum_{j=1}^N
(-1)^{\sum_{k=1}^{j-1}s_k}\id_{P(1)_{s_1}}\otimes\ldots\otimes p(j)_{s_j}\otimes\ldots\otimes\id_{P(N)_{s_N}}\,.\nonumber
\eeqn
Indeed, there is a slightly more elegant Koszul-type representation of this matrix factorisation (\cf Section 4.3 of \cite{Enger:2005jk}), which will be very useful later for the extraction of the large volume complex. In order to describe it, one introduces the vector space $V=S^N$ with basis $(e_i)_i$ and dual basis $(e_i^*)_i$. Then $P$ can be written as
\beq
P: \Lambda^{\rm odd} V\overset{\delta+\sigma}{\underset{\delta+\sigma}{\rightleftarrows}}\Lambda^{\rm even}V
\eeq
with\footnote{Note that $\delta\sigma+\sigma\delta=W\id$.}
\beq\label{eq:TPmaps} 
\delta=\imath_{\sum_j x_j e_j^*}\,,\quad
\sigma=(\sum_j A_j e_j)\wedge\cdot\,.
\eeq
As a side remark let us point out that the Koszul representation for tensor product matrix factorisation appears very naturally in the discussion of B-type boundary conditions in Landau-Ginzburg models. There, one typically introduces additional boundary fermions $\psi_i$ and their duals $\psi_i^*$ in order to achieve B-type supersymmetry despite the presence of the boundary \cite{Govindarajan:2001kr,Hellerman:2001bu,Govindarajan:2006uy}. These fermions satisfy a Clifford algebra with relations
 $\{\pi_i,\pi_j^*\}=\delta_{ij}$, and their presence gives rise to the boundary contribution 
\begin{equation}
     Q_{\rm bdry} = \sum_j \left( A_j \pi^*_j + x_j \pi_j \right)  
\end{equation}
of the supercharge. In the Koszul representation, these fermions are realised by the operators 
\beq
\pi_i=e_i\wedge\cdot\,,\quad
\pi_i^*=\imath_{e_i^*}\,,
\eeq
with corresponding Fock space $\Lambda^*V$. The supercharge can be easily recognised to be the matrix factorisation map
$Q_{\rm bdry}=\delta+\sigma$ in this representation. 

We are interested in the $\ZZ_d$-orbifold of the Landau-Ginzburg model with superpotential $W$, where $\ZZ_d$ acts on the $x_i$ by collective phase multiplications. The B-branes we would like to consider here are the ones associated to the matrix factorisations obtained from $P$ by means of the orbifold construction. Being invariant under $\ZZ_d$, the orbifold
just introduces $\ZZ_d$-representations on the modules $P_s$. The result are the following $\ZZ_d$-equivariant matrix factorisations specified by $m\in\ZZ_d$:
\beq
\wt P^m: \wt P_1^m\overset{\wt p_1^m}{\underset{\wt p_0^m}{\rightleftarrows}}\wt P_0^m\,,
\eeq
with 
\beq
\wt P^m_s=\mathop{\bigoplus_{(s_i)\in\ZZ_2^N}}_{s-\sum_i s_i\,{\rm even}}\left(\bigotimes_{i=1}^NP(i)_{s_i}\right)[m+\sum_is_i]
\eeq
and the maps $\wt p_s^m=p_s$. Here $[\cdot]$ denotes the $\ZZ_d$-representation on the respective modules. The Koszul-type representation can be made equivariant as well by setting the $\ZZ_d$-degree $[e_i]$ of the basis vectors $e_i$ of $V$ to be $1$. Then, $\wt P^m$ can be written as
\beq
\wt P^m: \left(\Lambda^{\rm odd} V\right)[m]\overset{\delta+\sigma}{\underset{\delta+\sigma}{\rightleftarrows}}\left(\Lambda^{\rm even}V\right)[m]
\eeq
with maps $\delta$ and $\sigma$ as above.

The ${\rm U}(1)_R$-representations can be specified by lifting $m$ to be an element of $\ZZ$. The
${\rm U}(1)_R$-charges of the modules $\bigotimes_i P(i)_{s_i}$ are then defined to be 
$\frac{2}{d}(m+\sum_i (s_i-d))$. In the Koszul-representation this amounts to giving $R$-charge $-1+\frac{2}{d}$
to the basis vectors $e_i$ and shifting the overall $R$-charge of the $\Lambda^s V$ by $\frac{2m}{d}$.

As discussed in Section~\ref{glsm} the Landau-Ginzburg orbifold can be obtained as a limit of a gauged linear $\sigma$-model with gauge group ${\rm U}(1)$, which in the LG-phase is broken down to $\ZZ_d$. Apart from the fields $x_j$ of charge $1$ which are also present in the LG-model, there is an additional chiral field $\pf$ of charge $-d$ in the GLSM. The $R$-charge of the fields $x_j$ and $\pf$ are $0$ and $2$ respectively, and the superpotential is given by $\widehat W(x_j,\pf)=\pf W(x_j)$. 

To transport B-branes from the LG-point to the large volume point of the
K\"ahler moduli space, we follow \cite{Herbst:2008jq} and first lift them to the GLSM.
Thus, as discussed in Section \ref{sec:GLSMtoLG} we need to construct ${\rm U}(1)\times{\rm U}(1)_R$-equivariant matrix factorisations which at the Landau-Ginzburg point reduce to the $\ZZ_d\times{\rm U}(1)_R$-equivariant matrix factorisations $\wt P^m$. Before doing that, let us first discuss the lifts of the tensor factors
\beq
\wt P(j)^m:S[m+1]\overset{\wt p(j)_1^{m}=x_j}{\underset{\wt p(j)_0^{m}=A_j}{\rightleftarrows}}S[m]
\eeq
to the GLSM. 
Each $\wt P(j)^m$ has two types of lifts:
\beq
\widehat P(j)^{[\alpha,a,r]}:\widehat S[a+1-d\alpha,r-1+2\alpha]\overset{\widehat p(j)_1^{[\alpha,a,r]}=\pf^{\alpha}x_j}{\underset{\widehat p(j)_0^{[\alpha,a,r]}=\pf^{1-\alpha} A_j}{\rightleftarrows}}\widehat S[a,r]\,,
\eeq
$m=a\,{\rm mod}\,d$, $\alpha\in\{0,1\}$ which differ in which of the maps $\wt p_s^m$ is multiplied by $\pf$. Here $\widehat S=\CC[x_i,\pf]$ and $[\cdot,\cdot]$ denotes the ${\rm U}(1)\times{\rm U}(1)_R$-representations. Moreover, the lifted $R$-charge is given by $r=2\frac{m-a}{d}$.

For the lift to the GLSM and the transport of the B-branes corresponding to $\wt P^m$ to the large volume phase, we restrict 
the discussion to the ``Calabi-Yau''-case, so from now on $d=N$. It is not difficult to see that in that case, there is only one lift 
of each $\wt P^m$ which is compatible with a given charge window $\NNN_{\eta}=\{1-d+\eta,\ldots,\eta\}$. This lift is given in the following way. First, choose $a=m-dk\in\NNN_\eta$ for some $k\in\ZZ$ and give ${\rm U}(1)\times{\rm U}(1)_R$-representation $[a,r=2k]$ to $\bigotimes_j P(j)_0$. Then, in the tensor product maps $\wt p_s^m=p_s$ written in \eq{tpmaps} one replaces the $p(j)_s$ by $\widehat p(j)_s^{[0,\cdot,\cdot]}$ everywhere except in the $(n=1+\eta-a)$th step
\beq
\mathop{\bigoplus_{(s_1,\ldots,s_N)}}_{\sum_i s_i=n} \bigotimes_j P(j)_{s_j}\rightleftarrows \mathop{\bigoplus_{(s_1,\ldots,s_N)}}_{\sum_i s_i=n-1} \bigotimes_j P(j)_{s_j}
\eeq
where one replaces them by $\widehat p(j)_s^{[1,\cdot,\cdot]}$. Noting that 
\beq
{\rm source}(\widehat p(j)_1^{[\alpha,a,r]})={\rm target}(\widehat p(j)_1^{[\beta,a+1-d\alpha,r-1+2\alpha]})
\eeq
one arrives at the following matrix factorisation
\beq
\widehat P^{[n,a,r]}: \widehat P^{[n,a,r]}_1\overset{\widehat p^{[n,a,r]}_1}{\underset{\widehat p^{[n,a,r]}_0}{\rightleftarrows}}
\widehat P^{[n,a,r]}_0\,,
\eeq
with 
\beq
\widehat P^{[n,a,r]}_s=\mathop{\bigoplus_{(s_i)\in\ZZ_2^N}}_{s-\sum_i s_i\,{\rm even}}\left(\bigotimes_{i=1}^NP(i)_{s_i}\right)[{\s a+\sum_is_i-d\Theta(\sum_i s_i-n),r-\sum_i s_i+2\Theta(\sum_i s_i-n)}]
\eeq
and 
\beqn
&&\widehat p_s^{[n,a,r]}=
\mathop{\sum_{(s_i)\in\ZZ_2^N}}_{s-\sum_i s_i\,{\rm even}}\sum_{j=1}^N
(-1)^{\sum_{k=1}^{j-1}s_k}\id_{P(1)_{s_1}}\otimes\ldots\\
&&\qquad\qquad\otimes
\widehat p(j)_{s_j}^{[\delta(S(j,s)-n),a+S(j,s)-d\Theta(S(j,s)-n),r-S(j,s)+2\Theta(S(j,s)-n)]}\otimes\ldots\otimes\id_{P(N)_{s_N}}\,,\nonumber
\eeqn
where $S(j,s)=\sum_{k\neq j}s_k$. 

In the Koszul-type representation this can be written as follows. First, set $\widehat V:=\widehat S^N$, and give degree 
$[1,-1]$ to its basis vectors $e_i$. Then $\widehat P^{[n,a,r]}$ can be written as
\beq
\widehat P^{[n,a,r]}: \bigoplus_{s\,{\rm odd}}\left(\Lambda^s \widehat V\right)[{\s a-d\Theta(s-n),r+2\Theta(s-n)}]
\overset{\widehat \delta+\widehat\sigma}{\underset{\widehat \delta+\widehat\sigma}{\rightleftarrows}}
\bigoplus_{s\,{\rm even}}\left(\Lambda^s \widehat V\right)[{\s a-d\Theta(s-n),r+2\Theta(s-n)}]\,,
\eeq
where 
\beq
\widehat\delta\big|_{\Lambda^s\widehat V}=\left\{\begin{array}{ll}
\delta\,,&{\rm for}\; s\neq n\\
\pf\delta\,,&{\rm for}\; s=n\end{array}\right.\qquad
\widehat\sigma\big|_{\Lambda^s\widehat V}=\left\{\begin{array}{ll}
\pf \sigma\,,&{\rm for}\; s\neq n-1\\
\sigma\,,&{\rm for}\; s=n-1\end{array}\right.\,.
\eeq
Let us transport the B-branes corresponding to $P^m$ from the Landau-Ginzburg to the large volume phase through the 
phase boundary component $N_0$.
The lift of $P^m$  compatible with the corresponding charge window $\NNN_0=\{1-d,\ldots,0\}$ 
is given by $\widehat P^{[1-a,a,r]}$,
where $a=m-dk\in\NNN_0$ for some $k\in\ZZ$, and $r=2k$. 
We now follow  \cite{Herbst:2008jq} to construct the 
B-brane obtained by transporting the LG-brane associated to $\wt P^m$ to the large volume phase 
along a path which traverses the phase boundary in the segment $N_0$. As described in Section \ref{sec:GLSMtoLV}
we first have to tensor the matrix factorisation $\widehat P^{[1-a,a,r]}$ by $\widehat S/(W)$. Then we regard it as a complex over $R:=\widehat S/(W,\pf)$ and ``unfold'' it with respect to ${\rm U}(1)_R$-charge. The sheafification of the resulting complex represents the large volume B-brane. For the case at hand, we use the Koszul-type representation of $\widehat P^{[1-a,a,r]}$ and introduce $V_R:=\widehat V\otimes R$. Then it is not difficult to see that the resulting complex indeed has the form
\beq
{\rm Cone}(\KKK^{-a}\stackrel{\sigma}{\rightarrow}\CCC)\{-r\}\,,
\eeq
where $\{\cdot\}$ denotes the shift of complexes and $\KKK^{-a}$ is the complex
\beq
\KKK^{-a}: \Lambda^{-a}V_R[a]\stackrel{\delta}{\rightarrow}\Lambda^{-a-1}V_R[a]\stackrel{\delta}{\rightarrow}\ldots \stackrel{\delta}{\rightarrow}\Lambda^0V_R[a]\rightarrow 0\,,
\eeq
starting at position $a$. Moreover, $\CCC$ is an infinite complex which itself can be written as successive cone:
\beq
\CCC={\rm Cone}(\KKK^{d}\{-2\}\otimes R[a]\stackrel{\sigma}{\rightarrow}\CCC\{-2\}\otimes R[-d])\,.
\eeq
Now $\KKK^{t}$ can be identified
as the dual of the Koszul complex of $(x_i)$ truncated at the $-(t+1)$st position. 
Its sheafification $\wt\KKK^{t}$ is exact except at the $-t$th position with cohomology $\wt M_{t}$. 
It is a well-known fact (\cf \eg Chapter 5B in \cite{Eisenbud2} -- for more details on the Koszul complex see \eg \cite{Eisenbud}) that the sheafification of $M_{t}$ is given by
\beq
\wt M_{t} \cong \Omega_{\PP^{N-1}}^{t}(t)\big|_X \,,
\eeq
where $\Omega^t_{\PP^{N-1}}(t)\big|_X$ is the vector bundle
$\Lambda^t\,T^*\PP^{N-1}\otimes\mathcal O(t)$ on the projective space
$\PP^{N-1}$ restricted to the hypersurface $X$.
It is zero for $t\geq N$. 
Thus, $\wt\KKK^{t}$ is quasi-isomorphic to 
\beq
\wt\KKK^{t}\cong \Omega_{\PP^{N-1}}^{t}(t)\big|_X \{t\}\,,
\eeq
whereas $\wt\KKK^d$ is quasi-isomorphic to the trivial complex. The latter implies that
$\CCC$ is quasi-isomorphic to the
trivial complex as well, and therefore, as summarised in Table~\ref{tab:TBBranes}, we obtain the large volume B-brane 
\beq
{\rm Cone}(\wt\KKK^{-a}\stackrel{\wt\sigma}{\rightarrow}\wt\CCC)\{-2k\}\cong\wt\KKK^{-a}\{-2k\}\cong\Omega_{\PP^{N-1}}^{-a}(-a)\big|_X\{-a-2k\}\,,
\eeq
by transporting the Landau-Ginzburg B-brane associated to the matrix factorisation $\wt P^m$, ($m=a+dk$, $a\in\NNN_0$, $k\in\ZZ$) into the large volume phase along 
a path intersecting the phase boundary in $N_0$. 
This is indeed the expected result \cite{Douglas:2000qw}.
\TABLE{
\begin{tabular}{|c|c|c|}
  \hline
  LG brane & GLSM brane & LV brane\vphantom{\rule[-1.5ex]{0pt}{4.8ex}} \\
  \hline\hline
  $\vdots$ & $\vdots$ & $\vdots$ \\
  \hline
  $\wt P^{0}$ & $\widehat P^{[1,0,0]}$ & 
    ${\Omega_{\PP^{N-1}}^0\big|_X(0)}\cong\underline{\mathcal O_X}$ 
  \vphantom{\rule[-1.5ex]{0pt}{4.8ex}} \\
  \hline
  $\wt P^{-1}$ & $\widehat P^{[2,-1,0]}$ & 
     $\Omega_{\PP^{N-1}}^1(1)\big|_X\{1\} 
       \cong \mathcal O_X^{\oplus d} \rightarrow \underline{\mathcal O_X(1)}$
  \vphantom{\rule[-1.5ex]{0pt}{4.8ex}}\\ 
  \hline
  $\wt P^{-2}$ & $\widehat P^{[3,-2,0]}$ & 
     $\Omega_{\PP^{N-1}}^2(2)\big|_X\{2\} 
       \cong \mathcal O_X^{\oplus\binom{d}{2}} \rightarrow \mathcal O_X(1)^{\oplus d}
       \rightarrow\underline{\mathcal O_X(2)}$ 
  \vphantom{\rule[-1.5ex]{0pt}{4.8ex}}\\
  \hline
  $\vdots$ & $\vdots$ & $\vdots$ \\
  \hline
  $\wt P^{-(d-1)}$ & $\widehat P^{[d,1-d,0]}$ & 
   {$\begin{array}{c}
     \Omega_{\PP^{N-1}}^{d-1}(d-1)\big|_X\{d-1\} \cong \\
     \mathcal O_X^{\oplus \binom{d}{d-1}} \rightarrow\cdots \rightarrow
         \mathcal O_X(d-2)^{\oplus \binom{d}{1}} \rightarrow\underline{\mathcal O_X(d-1)} 
     \end{array}$}
  \vphantom{\rule[-3.4ex]{0pt}{9ex}}\\
  \hline
  $\wt P^{-d}$ & $\widehat P^{[1,0,-2]}$ & 
    $\Omega_{\PP^{N-1}}^0\big|_X(0)\{2\}\cong{\mathcal O_X}\{2\}$ 
  \vphantom{\rule[-1.5ex]{0pt}{4.8ex}} \\
  \hline
  $\vdots$ & $\vdots$ & $\vdots$ \\
  \hline
\end{tabular}
\caption{\small The first column shows a collection of tensor product matrix factorisations describing B-branes 
in the Landau-Ginzburg phase. Their
lifts to the gauged linear $\sigma$-model compatible with charge window $\NNN_0$ are listed in the second column. The third column contains the large volume B-branes obtained by transporting the LG-branes through the phase boundary component $N_0$ into the large volume phase. Here underlining marks position zero in complexes.}\label{tab:TBBranes}}
\subsection{Defects} \label{defecttransport}
As discussed in Section \ref{defectmfsection}, at the Landau-Ginzburg point, defects can be described by
matrix factorisations $D$ of $W_1(x_1,\ldots,x_N)-W_2(y_1,\ldots,y_M)$. 
As such they can be transported to the large volume point in the same way as B-type D-branes. Namely, they can be lifted to defects between two GLSMs, \ie to matrix factorisations of $\pf W_1(x_1,\ldots,x_N)-\qf W(y_1,\ldots,y_M)$. This provides a uniform description of B-type defects on the K\"ahler moduli space of the product of the two GLSMs. 
The latter is the product of the two K\"ahler moduli spaces. In particular, one can transport B-type defects through the product $N\times N\p$ of phase boundary components of the respective models to the product of their large volume points. In this way, one obtains a complex ${\mathcal D}$ of coherent sheaves on the product $X\times Y$ of projective hypersurfaces $X=\{W_1=0\}\subset\PP^{N-1}$, $Y=\{W_2=0\}\subset\PP^{M-1}$ which represents the defect at large volume. We will now argue that this defect acts on the category $D^b(Y)$ as Fourier-Mukai transformation with kernel given by ${\mathcal D}\otimes(\OO_X\boxtimes \OO_Y(d)\{-1\})$.

Let us recall some basic facts about Fourier-Mukai transformations (for more details see \eg \cite{OFMT,BFMT}).
To every object ${\cal R}\in D^b(X\times Y)$ one can associate a functor
\beq
\Phi^{\mathcal R}:D^b(Y)\rightarrow D^b(X)\,,\qquad \Phi^{\cal R}(\,\cdot\,)={\rm\bf R}\pi_{1\,*}\left({\mathcal R}\stackrel{\rm\bf L}{\otimes}{\rm\bf L}\pi_2^*(\,\cdot\,)\right)\,,
\eeq
where $\pi_i$ denote the projections $X\stackrel{\pi_1}{\leftarrow}X\times Y\stackrel{\pi_2}{\rightarrow} Y$ on the factors, and `{\rm\bf L}' and `{\rm\bf R}' indicate that the tensor product, the  pullback and the pushforward are left- and right-derived respectively\footnote{For an introduction to derived categories and derived functors see \eg \cite{Gelfand:2002book}. In the following for ease of notation we do not explicitly denote in our formulae the left- or right-derived property of a functor.}.
In particular, for ${\mathcal F}\in D^b(X)$, ${\mathcal G}\in D^b(Y)$ this functor satisfies the following formula
\beq\label{FMHomformula}
\Hom({\mathcal F}\boxtimes {\mathcal G}^\lor,{\mathcal R})\cong\Hom({\mathcal F},\Phi^{\mathcal R}({\mathcal G}))\,.
\eeq
Here we used the abbreviation ${\mathcal F}\boxtimes{\mathcal G}=\pi_1^*({\mathcal F})\otimes\pi_2^*({\mathcal G})$.

Now, let us turn back to the Landau-Ginzburg point. Let $Q$ and $P$ be matrix factorisations of $W_1(x_i)$ and $W_2(y_i)$ respectively representing B-type D-branes in the corresponding Landau-Ginzburg models, and let $D$ be a matrix factorisation of $W_1(x_i)-W_2(y_i)$ describing a defect between these models.
Then the following ``folding formula'' holds\footnote{This is obvious from the folding trick, in which defects between two theories are regarded as boundary conditions in the tensor product of the first theory and the 
world-sheet parity dual of the second one.}
\beq\label{foldingformula}
\HH(Q,D*P)\cong\HH(Q\otimes P^*,D)\,,
\eeq 
where the dual $P^*$ of a matrix factorisation $P$ is defined by
\beq\label{eq:MFdual}
\left(P_1\overset{p_1}{\underset{p_0}{\rightleftarrows}}P_0\right)^*
=\left(P_1^*\overset{p_0^*}{\underset{-p_1^*}{\rightleftarrows}}P_0^*\right)\,.
\eeq
To transport these matrix factorisations to large volume, we fix components $N$ and $N\p$ of the phase boundary in the K\"ahler moduli space of the two models. Let us denote by ${\rm LV}_W$ the parallel transport from the Landau-Ginzburg point to large volume through the component $W$ of the phase boundary. Then define the complexes
\beqn
&&{\mathcal Q}:={\rm LV}_{N}(Q)\,,\;\;
{\mathcal P}:={\rm LV}_{N\p}(P)\,,\;\;
{\mathcal P}^*:={\rm LV}_{-N\p}(P^*)\,,\;\;
{\mathcal F}:={\rm LV}_{N}(D*P)\,,\\
&&
{\mathcal T}:={\rm LV}_{N\times(-N\p)}(Q\otimes P^*)\,,\;\;
{\mathcal D}:={\rm LV}_{N\times(-N\p)}(D)\,.\nonumber
\eeqn
Note that $-N\p$ is the component of the phase boundary dual to $N\p$, \ie the corresponding charge windows satisfy $\NNN\p=-\NNN$.

Since transport to large volume is an equivalence of categories \eq{foldingformula} implies
\beq
\Hom({\mathcal Q},{\mathcal F})\cong
\Hom({\mathcal T},{\mathcal D})\,.
\eeq
But it is not difficult to see that indeed 
\beq
{\mathcal T}\cong{\mathcal Q}\boxtimes{\mathcal P}^*\cong Q\boxtimes{\mathcal P}^{\lor}\otimes\OO_Y(-d)\{1\}\,,
\eeq
where $\mathcal{P}^\lor$ denotes the dual complex of $\mathcal{P}$. In the second step we used the relation
\beq
{\mathcal P}^*\cong{\mathcal P}^{\lor}\otimes\OO_Y(-d)\{1\}
\eeq
between duality of matrix factorisation and duality of large volume complexes
shown in Appendix \ref{dualcomplexes}.
Therefore, using \eq{FMHomformula} we arrive at
\beqn
\Hom({\mathcal Q},{\mathcal F})&\cong&
\Hom({\mathcal Q}\boxtimes\left({\mathcal P}\otimes\OO_Y(d)\{-1\}\right)^{\lor},{\mathcal D})\nonumber\\
&\cong&
\Hom({\mathcal Q},\Phi^{{\mathcal D}\otimes(\OO_X\boxtimes\OO_Y(d)\{-1\})}{\mathcal P})\,.
\eeqn
From this we deduce that fusion with the defect $D$ transported from the Landau-Ginzburg point to large volume through the component $N\times(-N\p)$ is realised by the Fourier-Mukai transform $\Phi^{\mathcal R}$ with kernel 
\beq\label{eq:KerDefRelation}
{\cal R}\cong{\mathcal D}\otimes(\OO_X\boxtimes\OO_Y(d)\{-1\})\,.
\eeq
\section{Landau-Ginzburg Monodromy}\label{LGmonsection}
The monodromies which are most easily described in the Landau-Ginzburg framework are those 
around the Landau-Ginzburg points themselves. These are orbifold points in K\"ahler moduli space, at which certain dualities become self-dualities. As a consequence, the symmetry groups of the theory at these points are enhanced by ``quantum symmetries'', which in turn realise the monodromies around these points. 

The models we are interested in here are one-parameter models which can be realised as gauged linear $\sigma$-models with ${\rm U}(1)$ gauge group, chiral superfields $(\pf,x_1,\ldots,x_N)$ of charge $(-d,1,\ldots,1)$ and superpotential $\pf W(x_1,\ldots,x_N)$, where $W\in\CC[x_1,\ldots,x_N]$ is homogeneous of degree $d$. At the Landau-Ginzburg point, the field $\pf$ receives a non-vanishing vacuum expectation value, and the model degenerates to a Landau-Ginzburg orbifold with chiral superfields $x_i$, superpotential $W$ and orbifold group $\ZZ_d$ to which the gauge group is broken. 

The action of the quantum symmetries in these orbifolds are well known. They act by phase multiplications in the twisted sectors. On the level of D-branes, this action is given by a shift in the representations of the orbifold group $\ZZ_d$ and the $R$-symmetry group ${\rm U}(1)_R$ \footnote{This is rather obvious from the gauged linear $\sigma$-model perspective \cite{Herbst:2008jq}, where boundary actions only depend on the $\theta$-angle through the combination $\theta+2\pi q$. Thus going around the Landau-Ginzburg point $\theta\mapsto\theta+2\pi$ shifts the ${\rm U}(1)$-representations $q$ of D-branes by $q\mapsto q-1$.}.

Using the results of \cite{Brunner:2007qu} one can easily construct defect matrix factorisations realising precisely this shift.
\subsection{Landau-Ginzburg Phase}\label{lglgmon}
In this Section we will construct the graded matrix factorisations representing the defects associated to Landau-Ginzburg monodromies. Let $W\in\CC[x_1,\ldots,x_N]$ be a homogeneous polynomial of degree $d$. We assume that the orbifold group $\Gamma=\ZZ_d$ acts on the $x_i$ by a collective phase multiplication. 

In the unorbifolded Landau-Ginzburg model, the identity defect 
is represented by the tensor product matrix factorisation (see \eg \cite{Brunner:2007qu})
\beq\label{tpid}
\II=\bigotimes_{i=1}^N\II(i)\,,
\eeq
of the rank-one matrix factorisations
\beq\label{N>1}
\II(i):\II(i)_1=S \overset{\imath(i)_1}{\underset{\imath(i)_0}{\rightleftarrows}}S=\II(i)_0\,,\;
\imath(i)_1=(x_i-y_i)\,,\; \imath(i)_0=A_i(x_j,y_j)\,.
\eeq
Here $S=\CC[x_j,y_j]$ and $A_i$ are homogeneous polynomials of degree $d-1$ such that\footnote{Although the $A_i$ are not determined uniquely by this equation, the equivalence class of $\II$ is independent of the choices of the $A_i$.} $W(x_j)-W(y_j)=\sum_{i=1}^N(x_i-y_i)A_i(x_j,y_j)$. 

The graded matrix factorisations corresponding to the Landau-Ginzburg monodromies in the $\Gamma$-orbifold can be obtained from $\II$ by means of the orbifold construction. For the case of Landau-Ginzburg models with one chiral field, \ie $N=1$ this has been discussed in \cite{Brunner:2007ur}. Let us briefly review this construction for the rank-one factorisations \eq{N>1}. Since the defects
separate two orbifold theories, they have to be orbifolded by the product of the orbifold groups on either side. In this case the group is 
$\widehat{\Gamma}=\Gamma\times\Gamma=\ZZ_d\times\ZZ_d$, where the first factor acts on the variables $x_j$ and the second one on the variables $y_j$ only. As usual, the orbifold procedure requires to choose on the given unorbifolded matrix factorisation a representation of the subgroup $\widehat{\Gamma}_s\subset\widehat{\Gamma}$ which stabilises it. Then one sums over its orbit  
under $\widehat{\Gamma}/\widehat{\Gamma}_s$. The result is a $\widehat{\Gamma}$-equivariant matrix factorisation. 

The matrix factorisations $\II(i)$ are invariant under the diagonal subgroup $\Gamma_s=\Gamma_{\rm diag}\subset\Gamma\times\Gamma$, because the maps $\imath(i)_s$ are homogeneous polynomials. Therefore one has to choose a representation of $\Gamma\cong\Gamma_s$ on it, which we will denote by $m\in\ZZ_d$. 
The next step is to sum up the $\Gamma\times\Gamma/\Gamma_{\rm diag}\cong\Gamma$ orbit. The result is a direct sum of $d$ rank-one factorisations. By construction the action of $\Gamma\times\Gamma$ is not diagonal in the sum basis, but one can diagonalise it. 
In the corresponding basis the resulting $\Gamma\times\Gamma$-equivariant matrix factorisations $\wt{\II}(i)^m$ are given by
\beqn\label{idmf}
&&\wt\II(i)^m:
\wt\II(i)^m_1=V^{m+1}\overset{\wt\imath(i)_1^m}{\underset{\wt\imath(i)_0^m}{\rightleftarrows}}V^m=\wt\II(i)^m_0\,,\\
&&{\rm where}\quad V^m=\bigoplus_{\nu=0}^{d-1} S[m+\nu,-\nu]\\
&&
\wt\imath(i)_1^m=
\left(\begin{array}{cccc} \s x_i & & & \s -y_i \\ \s -y_i & \ddots & & \\ &
\s\ddots &\s\ddots & \\&& \s-y_i&\s x_i\end{array}\right)\,,\qquad
(\wt\imath(i)_0^m)_{\mu\nu}=A_i^{\mu-\nu}(x_j,y_j)\,.\nonumber
\eeqn
Here $[\cdot,\cdot]$ specifies the $\ZZ_d\times\ZZ_d$-representation on this module, and 
$A_i^l(x_j,y_j)$ are the degree $[d-1-l,l]$-parts of the polynomials $A_i(x_j,y_j)$. To write this in a more compact form,
denote by $(f_\mu^m)_\mu$ the basis of $V^m$ with degrees
$[f^m_\mu]=[m+\mu,-\mu]$, and define the map $\tau:V^m\rightarrow V^m$, $f_\mu^m\mapsto f_{\mu+1\,{\rm mod}\,d}^m$.
Then
\beq
\wt\imath(i)_1^m=x_i\id-y_i\tau\,,\quad\wt\imath(i)_0^m=\sum_{l=0}^{d-1}A_i^l(x_j,y_j)\tau^l\,.
\eeq
For a more detailed discussion see \cite{Brunner:2007ur}.

Indeed, this construction can easily be generalised to the matrix factorisation \eq{tpid} corresponding to the identity defect.
As tensor product the latter has the following form:
\beq
\II:\II_1\overset{\imath_1}{\underset{\imath_0}{\rightleftarrows}}\II_0\,,
\eeq
where 
\beq
\II_s
=\mathop{\bigoplus_{(s_i)\in\ZZ_2^N}}_{s-\sum_i s_i\,{\rm even}}\bigotimes_{i=1}^N\II(i)_{s_i}\,,
\eeq
and
\beq
\imath_s=
\mathop{\sum_{(s_i)\in\ZZ_2^N}}_{s-\sum_i s_i\,{\rm even}}\sum_{j=1}^N
(-1)^{\sum_{k=1}^{j-1}s_k}\id_{\II(1)_{s_1}}\otimes\ldots\otimes\imath(j)_{s_j}\otimes\ldots\otimes\id_{\II(N)_{s_N}}\,.
\eeq
As in the case of tensor product B-branes discussed in Section \ref{tpbranes}, a Koszul-type representation
of this matrix factorisation is useful.
Let $V:=S^N$ with basis $(e_i)_i$ and denote the dual basis by $(e_i^*)_i$. 
Then the identity matrix factorisation \eq{tpid} can be represented by
\beq\label{koszulrepresentation}
\II: \Lambda^{\rm odd} V\overset{\delta+\sigma}{\underset{\delta+\sigma}{\rightleftarrows}}\Lambda^{\rm even}V\,,
\eeq
with\footnote{As in the case of the tensor product D-branes discussed in Section \ref{tpbranes} $\delta\sigma+\sigma\delta=(W(x_j)-W(y_j))\id$.}
\beq
\delta=\imath_{\sum_i(x_i-y_i)e_i^*}\,,\quad \sigma=(\sum_i A_ie_i)\wedge\cdot\,.
\eeq
Applying the orbifold construction to this matrix factorisation yields the equivariant matrix factorisations 
\beq
\wt{\II}^m:\wt{\II}^m_1\overset{\wt\imath_1^m}{\underset{\wt\imath_0^m}{\rightleftarrows}}\wt{\II}^m_0
\eeq
with
\beq\label{imodules}
\wt{\II}^m_s
=\mathop{\bigoplus_{(s_i)\in\ZZ_2^N}}_{s-\sum_i s_i\,{\rm even}}\bigotimes_{i=1}^N\II(i)_{s_i}\otimes V^{m+\sum_is_i}\,,
\eeq
and
\beq\label{tpequiv}
\wt\imath_s^m=
\mathop{\sum_{(s_i)\in\ZZ_2^N}}_{s-\sum_i s_i\,{\rm even}}\sum_{j=1}^N
(-1)^{\sum_{k=1}^{j-1}s_k}\id_{\II(1)_{s_1}}\otimes\ldots\otimes\wt\imath(j)_{s_j}^{m+\sum_{k\neq j} s_k}\otimes\ldots\otimes\id_{\II(N)_{s_N}}\,.
\eeq
To describe the result of the orbifold construction in the Koszul-representation \eq{koszulrepresentation}, we set 
the $\ZZ_d\times\ZZ_d$-degree of the basis vectors $e_i$ of $V$ to $[1,0]$.
Then the orbifolds of the identity matrix factorisation can be written as 
\beq\label{koszulrep}
\wt\II^m:\Lambda^{\rm odd}V\otimes V^m\overset{\wt\delta+\wt\sigma}{\underset{\wt\delta+\wt\sigma}{\rightleftarrows}}\Lambda^{\rm even}V\otimes V^m\,,
\eeq
with
\beq\label{eq:LGmaps} 
\wt\delta=\imath_{\sum_i x_ie_i^*}\otimes{\rm id}-\imath_{\sum_iy_ie_i^*}\otimes\tau\,,\quad
\wt\sigma=\sum_{i,l}A_i^l(e_i\wedge\cdot)\otimes\tau^l\,.
\eeq
The ${\rm U}(1)_R$-representations on the matrix factorisations $\wt\II^m$ are specified as follows. Similarly to the case of tensor product D-branes discussed in Section \ref{tpbranes}, $m$ is lifted to an integer, and the $R$-charge of vectors in $V^{m+\sum_i s_i}$ in \eq{imodules} is defined to be $\frac{2}{d}(m+\sum_i(s_i-d))$. In the Koszul-representation \eq{koszulrep} one also sets the $R$-charge of vectors of $V^m$ to $\frac{2m}{d}$, and the ones of the basis vectors $e_i$ of $V$ to $-1+\frac{2}{d}$.
\subsection{Lift to the GLSM}\label{glsmlift}
Next, we will lift 
the matrix factorisations corresponding to Landau-Ginzburg monodromies which have been constructed in 
Section \ref{lglgmon} to the gauged linear $\sigma$-model. Apart from the chiral fields $x_i$ which were also present in the Landau-Ginzburg model,
this model 
contains an additional field $\pf$, 
and the gauge group which is broken to $\Gamma=\ZZ_d$ in the Landau-Ginzburg phase is ${\rm U}(1)$. The $x_i$ have charge $1$ with respect to it, and 
the field $\pf$ charge $-d$. The model also exhibits a ${\rm U}(1)_R$, with charges $0$ and $2$ of the fields $x_i$ and $\pf$ respectively.
The superpotential of the model is given by $\widehat{W}(\pf,x_j)=\pf W(x_j)$. To lift a $\ZZ_d\times{\rm U}(1)_R$-equivariant matrix factorisation $P$ of $W$ 
from the Landau-Ginzburg to the gauged linear $\sigma$-model means to construct ${\rm U}(1)\times{\rm U}(1)_R$-equivariant
matrix factorisations of $\widehat{W}$ which in the Landau-Ginzburg phase reduce to $P$. As alluded to in Section \ref{sec:GLSMtoLG} there
are in general several lifts of a given matrix factorisation $P$. We are going to lift the defect matrix factorisations $\wt\II^m$ constructed in Section \ref{lglgmon}, \ie a $\ZZ_d\times\ZZ_d\times{\rm U}(1)_R$-equivariant matrix factorisation of $W(x_j)-W(y_j)$ to ${\rm U}(1)\times{\rm U}(1)\times{\rm U}(1)_R$-equivariant matrix factorisations of $\widehat{W}(\pf,x_j)-\widehat{W}(\qf,y_j)$. In order to do so, we will again start by lifting the orbifold factorisations $\wt\II(i)^m$ of the tensor factors of the identity defect in the unorbifolded Landau-Ginzburg models. 

It is not difficult to see how to promote the factorisations $\wt\II(i)^m$ to the gauged linear $\sigma$-model. To be able to lift
the $\ZZ_d\times\ZZ_d$-representations to ${\rm U}(1)\times{\rm U}(1)$-representations, one needs to
replace one of the entries $x_i$ of $\wt\imath(i)^m_1$ by $\pf x_i$, and one of the entries $y_i$ by $\qf y_i$. 
At the same time, one has to 
introduce monomial factors in $\pf$ and $\qf$ in the entries of $\wt\imath(i)^m_0$ accordingly.
Once the factors of $\pf$ and $\qf$ have been placed in $\wt\imath(i)^m_1$ the distribution of factors 
in $\wt\imath(i)^m_0$ is fixed. The lifts of the matrix factorisations $\wt\II(i)^m$ are given by 
\beq\label{liftfactor}
\widehat\II(i)^{[n,a,b,r]}:
V^{[n+1,a+1,b,r-1]}
\overset{\widehat\imath(i)_1^{[n,a,b,r]}}{\underset{\wt\imath(i)_0^{[n,a,b,r]}}{\rightleftarrows}}
V^{[n,a,b,r]}\,,
\eeq
with
\beq
V^{[n,a,b,r]}=\bigoplus_{\nu=0}^{d-1}\widehat S[a+\nu-d\Theta(\nu-d+n),b-\nu,r+2\Theta(\nu-d+n)]\,,
\eeq
and
\beq
\widehat\imath(i)_1^{[n,a,b,r]}=
\left(\begin{array}{ccccc} 
\s x_i & & & & \s -\qf y_i \\ 
\s -y_i & \ddots & & & \\ 
& \s\ddots &\s\pf x_i & & \\
& & \s\ddots&\s\ddots &\\
& & & \s-y_i&\s x_i\end{array}\right)\,,\;
\left(\widehat\imath(i)_0^{[n,a,b,r]}\right)_{\mu\nu}=Z^{\mu,\nu}(\pf,\qf)A_i^{\mu-\nu}(x_j,y_j)\,.
\eeq
Here $\widehat S=\CC[x_j,y_j,\pf,\qf]$, $[\cdot,\cdot,\cdot]$ specifies the ${\rm U}(1)\times{\rm U}(1)\times{\rm U}(1)_R$-representation
and the entry $\pf x_i$ in $\widehat\imath(i)^{[n,a,b,r]}_1$ is at the $(d-n)$th position.
The matrix $Z^{\mu\nu}$ has the form
\beq
Z(\pf,\qf)=\left(\begin{array}{cccc|c|cccc}
\pf & \qf & \ldots & \qf & \qf &\pf\qf &\ldots&\ldots&\pf\qf \\
\vdots & \ddots & \ddots &\vdots&\vdots & \vdots&\ddots&&\vdots\\
\vdots & & \ddots & \qf & \vdots & \vdots &  & \ddots & \vdots\\
\pf & \ldots &\ldots &\pf & \qf& \pf\qf&\ldots&\ldots&\pf\qf\\
\hline
1 &\ldots&\ldots&1&1&\qf&\ldots&\ldots&\qf\\
\hline
1 &\ldots&\ldots&1&1 &\pf&\qf&\ldots&\qf\\
\vdots&\ddots&&\vdots&\vdots&\vdots&\ddots&\ddots&\vdots\\
\vdots&&\ddots&\vdots&\vdots&\vdots&&\ddots &\qf\\
1&\ldots&\ldots&1&1&\pf&\ldots&\ldots&\pf
\end{array}\right)\,,
\eeq
where the $(d-n)$th row and column have been marked. 
Omitting factors of $\pf$ and $\qf$ which can be reconstructed out of the ${\rm U}(1)\times{\rm U}(1)\times{\rm U}(1)_R$-representations by noting that the maps $\widehat\imath(i)^{[n,a,b,r]}_s$ have degree $[0,0,1]$, this can again be written in a more compact form. 
Let $(f_\mu^{[n,a,b,r]})_\mu$ be a basis\footnote{the basis used to write \eq{liftfactor}} of $V^{[n,a,b,r]}$ with $[f_\mu^{[n,a,b,r]}]=[a+\mu-d\Theta(\mu-d-n),b-\mu,r+2\Theta(\mu-d-n)]$ and define $\widehat{\tau}:f_\mu^{[n,a,b,r]}\mapsto f_{\mu+1\,{\rm mod}\,d}^{[n,a,b,r]}$, then  
\beq
\widehat\imath(i)^{[n,a,b,r]}_1=x_i\id-y_i\widehat\tau\,,\quad 
\widehat\imath(i)^{[n,a,b,r]}_0=\sum_{l=0}^{d-1}A_i^l(x_j,y_j)\widehat\tau^l\,.
\eeq
The GLSM matrix factorisations $\widehat\II(i)^{[n,a,b,r]}$ are labelled by representations $(a,b,r)$ of ${\rm U}(1)\times{\rm U}(1)\times{\rm U}(1)_R$ and $n\in\{0,\ldots,d-1\}$, and it makes sense to define 
\beq
\widehat\II^{[n+dk,a,b,r]}:=\widehat\II^{[n,a-dk,b,r+2k]}\,.
\eeq
It is easy to see that $\widehat\II(i)^{[n,a,b,r]}$ reduces to the matrix factorisation $\wt\II(i)^{m}$ at the Landau-Ginzburg point if 
\beq
(a+b-m)\;{\rm mod}\;d=0\quad {\rm and}\quad r=\frac{2}{d}(m-a-b)\,.
\eeq
Given these lifts, one can easily lift the orbifold matrix factorisation $\wt\II^m$ (\cf \eq{tpequiv}) 
of the identity factorisation $\II$ to the GLSM. Using the fact
\beq
{\rm source}\left(\widehat\imath(i)^{[n,a,b,r]}_1\right)=
{\rm target}\left(\widehat\imath(i)^{[n+1,a+1,b,r-1]}_1\right)
\eeq
one arrives at
\beq
\widehat\II^{[n,a,b,r]}:\widehat\II^{[n,a,b,r]}_s
\overset{\widehat\imath_1^{[n,a,b,r]}}{\underset{\widehat\imath_0^{[n,a,b,r]}}{\rightleftarrows}}
\widehat\II^{[n,a,b,r]}_s\,,
\eeq
with
\beq
\widehat{\II}^{[n,a,b,r]}_s
=\mathop{\bigoplus_{(s_i)\in\ZZ_2^N}}_{s-\sum_i s_i\,{\rm even}}\bigotimes_{i=1}^N\left(\II(i)_{s_i}\otimes\widehat{S}\right)\otimes V^{[n+\sum_is_i,a+\sum_i s_i,b,r-\sum_is_i]}\,,
\eeq
and
\beqn\label{glsmtpequiv}
&&\widehat\imath_s^{[n,a,b,r]}=
\mathop{\sum_{(s_i)\in\ZZ_2^N}}_{s-\sum_i s_i\,{\rm even}}\sum_{j=1}^N
(-1)^{\sum_{k=1}^{j-1}s_k}\times\\
&&\qquad\qquad\times\id_{\II(1)_{s_1}\otimes\widehat S}\otimes\ldots\otimes\widehat\imath(j)_{s_j}^{[n+\sum_{k\neq j} s_k,a+\sum_{k\neq j} s_k,b,r-\sum_{k\neq j} s_k]}\otimes\ldots\otimes\id_{\II(N)_{s_N}\otimes\widehat S}\,,\nonumber
\eeqn
where as before $(a+b-m)\;{\rm mod}\;d=0$ and $r=\frac{2}{d}(m-a-b)$.
Again this can be more elegantly formulated in a Koszul-type representation. For this define $\widehat{V}:=\widehat{S}^N$ with basis $(e_i)_i$ of ${\rm U}(1)\times{\rm U}(1)\times{\rm U}(1)_R$-degree $[1,0,-1]$. Then the matrix factorisations 
$\widehat{\II}^{[n,a,b,r]}$ can be represented as
\beq
\widehat{\II}^{[n,a,b,r]}:
\bigoplus_{s\,{\rm odd}}\Lambda^s\widehat{V}\otimes V^{[s+n,a,b,r]}
\overset{\widehat\delta+\widehat\sigma}{\underset{\widehat\delta+\widehat\sigma}{\rightleftarrows}}
\bigoplus_{s\,{\rm even}}\Lambda^s \widehat V\otimes V^{[s+n,a,b,r]}\,,
\eeq
with
\beq
\widehat\delta=\imath_{\sum_i x_ie_i^*}\otimes{\rm id}-\imath_{\sum_iy_ie_i^*}\otimes\widehat\tau\,,\quad
\widehat\sigma=\sum_{i,l}A_i^l(e_i\wedge\cdot)\otimes\widehat\tau^l\,.
\eeq
Here, for ease of notation, factors of $\pf$, $\qf$ have been omitted. Their positions can be reconstructed out of the ${\rm U}(1)\times{\rm U}(1)\times{\rm U}(1)_R$-degrees by noting that the maps $\widehat\delta+\widehat\sigma$ have degree $[0,0,1]$.

The matrix factorisation $\widehat\II^{[n,a,b,r]}$ has charges in the window $\NNN_{a+d-n-1}\times\NNN_{b}$. Thus, the lift of the matrix factorisation
$\wt\II^m$ compatible with a window $\NNN_{\eta_1}\times\NNN_{\eta_2}$ is given by $\widehat\II^{[n,n+\eta_1+1-d,\eta_2,r]}$ for $n\in\{0,\ldots,d-1\}$ such
that $n=m-1-\eta_1-\eta_2\,{\rm mod}\,d$. 
\subsection{Transport to the Large Volume Phase}\label{lgmonodromylv}
Having lifted the matrix factorisations $\wt\II^m$ representing the LG monodromies to the gauged linear $\sigma$-model, we can now transport them to large volume to obtain the corresponding Fourier-Mukai kernels as discussed in Section \ref{defecttransport}. Using the Koszul-type representation
one easily sees that the large volume complex obtained by transporting the GLSM defect matrix factorisation $\widehat\II^{[n,a,b,r]}$ from
the Landau-Ginzburg point through phase boundary component $N_{a+d-n-1}\times N_{b}$ into the large volume phase
has the form 
\beq
{\rm Cone}(\wt\KKK_0^{[n,a,b,r]}\stackrel{\wt\sigma}{\rightarrow}\wt\CCC_0^{[n,a,b,r]})\,.
\eeq
Here
\beqn
\KKK_0^{[n,a,b,r]}:&& \Lambda^{d-n-1} V\otimes V^{[d-1,a,b,r-d+n+1]}_0\stackrel{\delta}{\rightarrow}
\Lambda^{d-n-2} V\otimes V^{[d-2,a,b,r-d+n+2]}_1\stackrel{\delta}{\rightarrow}\ldots\nonumber\\
&&\ldots\stackrel{\delta}{\rightarrow}
\Lambda^0 V\otimes V^{[n,a,b,r]}_{d-1}\,,
\eeqn
where the ``unhatted'' $V$'s are defined by $V=\widehat V\otimes_{\widehat S}\widehat S/(\pf,\qf,W(x_i),W(y_i))$. 
For $\mu=d-n-1$,
$V^{[n,a,b,r]}_\mu={\rm spann}_{R}(f_\nu^{[n,a,b,r]})_{0\leq\nu\leq\mu}$ are the subspaces of $\widehat V^{[n,a,b,r]}$ of
${\rm U}(1)_R$-charge $r$.
In more detail this complex looks like
\beq\label{triang}{\scriptsize
\begin{array}{ccccccc}
\Lambda^{d-n-1} V[a,b] & \rightarrow & \Lambda^{d-n-2} V[a,b] & \rightarrow & \ldots & \rightarrow & \Lambda^0 V[a,b]\\
&\searrow&\oplus&\searrow&\ldots&\searrow & \oplus\\
& & \Lambda^{d-n-2} V[a+1,b-1] & \rightarrow & \ldots & \rightarrow & \Lambda^0 V[a+1,b-1]\\
&&&\searrow&\ldots&\searrow & \oplus\\
&&&&&\ldots&\vdots\\
&&&&&&\Lambda^0 V[a+d-n-1,b-d+n+1]
\end{array}}\,,
\eeq
where the $U(1)_R$-degree has been dropped. Horizontal and diagonal arrows represent the maps $\imath_{\sum_j x_je_j^*}$ and $\imath_{\sum_j y_j e_j^*}$ respectively. As discussed in Section \ref{tpbranes} the sheafifications of the rows in this complex are quasi-isomorphic to 
\beq
\Omega_{\PP^{N-1}}^{d-n-1-i}(-a-i)\big|_X\boxtimes\OO_X(-b+i)\{-r+d-n-1-i\}\,,
\eeq
where $X$ is the projective hypersurface $\{W=0\}$ in $\PP^{N-1}$. 
The sheafification $\wt\KKK_0^{[n,a,b,r]}$ can therefore be written as 
\beqn\label{K0seq}
&&{\scriptsize \Omega_{\PP^{N-1}}^{d-n-1}(-a)\big|_X\boxtimes\OO_{X}(-b)\rightarrow 
\Omega_{\PP^{N-1}}^{d-n-2}(-a-1)\big|_X\boxtimes\OO_X(-b+1)\rightarrow\ldots\nonumber}\\
&&{\scriptsize \qquad\qquad\qquad\ldots\rightarrow
\Omega_{\PP^{N-1}}^{0}(-a-d+n+1)\big|_X\boxtimes\OO_X(-b+d-n-1)\,,}
\eeqn
where the last term is at position $r$. 

To discuss the complexes $\CCC_0^{[n,a,b,r]}$, define $V^{[n,a,b,r]}_\mu$ for all $\mu$ to be the subspace of $\widehat{V}^{[n,a,b,r]}$ of ${\rm U}(1)_R$-charge $r+2((\mu+n)\,{\rm div}\,d)$, and denote its basis vectors by
$f^{[n,a,b,r]}_\nu=\pf^{(\mu+n)\,{\rm div}\,d-(\nu+n)\,{\rm div}\,d}\qf^{\nu\,{\rm div}\, d}f^{[n,a,b,r]}_{\nu\,{\rm mod}\,d}$. 

Indeed, as in the discussion of the tensor product B-branes in Section \ref{tpbranes} the complex $\CCC_0^{[n,a,b,r]}$ can be written as 
successive cone:
\beq
\CCC_i^{[n,a,b,r]}={\rm Cone}(\KKK_{i+1}^{[n,a,b,r]}\stackrel{\sigma}{\rightarrow}\CCC_{i+1}^{[n,a,b,r]})\,,
\eeq
where 
\beq
\KKK_{i+1}^{[n,a,b,r]}=(\pf+\qf)\KKK_i^{[n,a,b,r]}\quad{\rm and}\quad
\CCC_{i+1}^{[n,a,b,r]}=(\pf+\qf)\CCC_i^{[n,a,b,r]}\quad{\rm for}\; i>0\,,
\eeq
and $\CCC_{i}^{[n,a,b,r]}$ is an infinite complex which starts at position $r-d+2+2i$.
Moreover
\beqn
\KKK_1^{[n,a,b,r]}:&&
\Lambda^{d} V\otimes V^{[n+d,a,b,r-d]}_{d-n}\stackrel{\delta}{\rightarrow}
\Lambda^{d-1} V\otimes V^{[n+d-1,a,b,r-d+1]}_{d-n+1}\stackrel{\delta}{\rightarrow}\ldots\nonumber\\
&&\ldots\stackrel{\delta}{\rightarrow}\Lambda^{i} V\otimes V^{[n+i,a,b,r-i]}_{2d-i-n}\stackrel{\delta}{\rightarrow}
\Lambda^0 V\otimes V^{[n,a,b,r]}_{2d-n}\,,
\eeqn
Similarly to $\KKK_0^{[n,a,b,r]}$, also $\KKK_1^{[n,a,b,r]}$ can be written in triangular shape \eq{triang}, 
and its sheafification $\wt\KKK_1^{[n,a,b,r]}$ is quasi-isomorphic to
\beqn\label{complex}
&&{\scriptsize \Omega_{\PP^{N-1}}^{d-1}(-a+n)\big|_X\boxtimes\OO_{X}(-b+d-n)\rightarrow}\\
&&{\scriptsize \qquad\rightarrow
\Omega_{\PP^{N-1}}^{d-2}(-a+n-1)\big|_X\boxtimes\OO_X(-b+d-n+1)\rightarrow\ldots\nonumber}\\
&&{\scriptsize \qquad\qquad\qquad\ldots\rightarrow
\Omega_{\PP^{N-1}}^{0}(-a+n-d+1)\big|_X\boxtimes\OO_X(-b+2d-n-1)\,,}\nonumber
\eeqn
where the last term is at position $r+2$. Indeed, $\wt\KKK_i^{[n,a,b,r]}$ for any $i\geq 0$ is quasi-isomorphic to the complex
\eq{complex} tensored by $\OO_{X}\boxtimes\OO_{X}(di)$ and shifted to the right by $2i$.

Using Beilinson's resolution 
\beqn
&&0\rightarrow\Omega^{N-1}_{\PP^{N-1}}(N-1)\boxtimes\OO_{\PP^{N-1}}(1-N)\rightarrow\ldots\rightarrow
\Omega^{1}_{\PP^{N-1}}(1)\boxtimes\OO_{\PP^{N-1}}(-1)\\
&&\qquad\qquad\qquad\qquad\qquad\qquad\qquad\qquad
\rightarrow\Omega^{0}_{\PP^{N-1}}(0)\boxtimes\OO_{\PP^{N-1}}(0)\rightarrow
\OO_{\Delta}\rightarrow 0\nonumber
\eeqn
of the diagonal $\Delta\subset\PP^{N-1}\times\PP^{N-1}$, we easily see that
\beq
\wt\KKK_i^{[n,a,b,r]}\cong\OO_{\Delta}\otimes\left(\OO_X(-a+n-d+1)\boxtimes\OO_X(-b+d-n-1+di)\right)\{-r-2i\}\,.
\eeq
 for $i>0$. One can then use the resolution
 \beqn
&&0\rightarrow\OO_{\PP^{N-1}}(-d)\boxtimes\OO_{\PP^{N-1}}(-d)\rightarrow(\OO_{\PP^{N-1}}(-d)\boxtimes\OO_{\PP^{N-1}})\oplus(\OO_{\PP^{N-1}}\boxtimes\OO_{\PP^{N-1}}(-d))\nonumber\\
&&\qquad\qquad\qquad\qquad\qquad\qquad\qquad\qquad\rightarrow\OO_{\PP^{N-1}}\boxtimes\OO_{\PP^{N-1}}\rightarrow \OO_{X\times X}\rightarrow 0
 \eeqn
 of $\OO_X\boxtimes\OO_X$ on $\PP^{N-1}\times\PP^{N-1}$ to obtain
 \beq\label{KKKcohomology}
 \wt\KKK_i^{[n,a,b,r]}\cong\OO_{\Delta X}(-a-b+(i-1)d)\{-r-2i+1\}\oplus\OO_{\Delta X}(-a-b+di)\{-r-2i\}
 \eeq
 for $i>0$. In this complex all maps are zero, \ie $\wt\KKK_i^{[n,a,b,r]}$ is quasi-isomorphic to its cohomology. Now, it is not difficult to see that the map $\sigma:\wt\KKK_i^{[n,a,b,r]}\rightarrow\wt\KKK_{i+1}^{[n,a,b,r]}$ in cohomology descends to the map
 $1:\OO_{\Delta X}(-a+b+di)\{-r-2i\}\rightarrow\OO_{\Delta X}(-a+b+di)\{-r-2i\}$ from the second summand in \eq{KKKcohomology} for $\wt\KKK_i^{[n,a,b,r]}$ to the first summand for $\wt\KKK_{i+1}^{[n,a,b,r]}$. Thus these summands successively cancel in the cone construction of $\wt\CCC_0^{[n,a,b,r]}$, and one ends up with
 \beq
 \wt\CCC_0^{[n,a,b,r]}\cong\OO_{\Delta X}(-a-b)\{-r-1\}\,.
 \eeq
Thus, this indeed completes \eq{K0seq} to the large volume kernel
\beqn\label{kernel}
&&{\scriptsize \Omega_{\PP^{N-1}}^{d-n-1}(-a)\big|_X\boxtimes\OO_{X}(-b)\rightarrow 
\Omega_{\PP^{N-1}}^{d-n-2}(-a-1)\big|_X\boxtimes\OO_X(-b+1)\rightarrow\ldots}\\
&&{\scriptsize \qquad\qquad\qquad\ldots\rightarrow
\Omega_{\PP^{N-1}}^{0}(-a-d+n+1)\big|_X\boxtimes\OO_X(-b+d-n-1)\rightarrow\OO_{\Delta X}(-a-b)}\,,
\nonumber
\eeqn
where the last term is at position $r+1$. 

Let us transport the LG-monodromy defect matrix factorisations $\wt\II^m$ from the Landau-Ginzburg point through the phase boundary component 
$N_0\times N_{d-1}$ corresponding to the combination of dual charge windows $\NNN_0\times(-\NNN_0)$ into the large volume phase. 
To simplify notation we parametrise $m=M+dk$ with $M\in\{-d,\ldots,-1\}$.
As discussed at the end of Section \ref{glsmlift}, 
the lift of the matrix factorisation $\wt\II^m$ to this window is given by $\widehat{\II}^{[M+d,M+1,d-1,2k-2]}$. Following the arguments above, the corresponding large volume complex can be read off from \eq{kernel} to be
\beqn\label{complm}
&&\left({\small \Omega_{\PP^{N-1}}^{-M-1}(-M-1)\big|_X\boxtimes\OO_{X}(1-d)\rightarrow 
\Omega_{\PP^{N-1}}^{-M-2}(-M-2)\big|_X\boxtimes\OO_X(2-d)\rightarrow\ldots}\right.\nonumber\\
&&\left.{\scriptsize \qquad\qquad\qquad\ldots\rightarrow
\Omega_{\PP^{N-1}}^{0}(0)\big|_X\boxtimes\OO_X(-M-d)\rightarrow\OO_{\Delta X}(-M-d)}\right)\{1-2k\}\,.
\eeqn
As discussed in Section \ref{glsm} the associated Fourier-Mukai kernels are obtained by tensoring these complexes by
$\OO_X\boxtimes\OO_X(d)\{-1\}$:
\beqn\label{kernelm}
&&\left({\scriptsize \Omega_{\PP^{N-1}}^{-M-1}(-M-1)\big|_X\boxtimes\OO_{X}(1)\rightarrow 
\Omega_{\PP^{N-1}}^{-M-2}(-M-2)\big|_X\boxtimes\OO_X(2)\rightarrow\ldots}\right.\nonumber\\
&&\left.{\scriptsize \qquad\qquad\qquad\ldots\rightarrow
\Omega_{\PP^{N-1}}^{0}(0)\big|_X\boxtimes\OO_X(-M)\rightarrow\OO_{\Delta X}(-M)}\right)\{-2k\}\,.
\eeqn
We indeed find the expected (see \cite{Aspinwall:2001dz,Aspinwall:2002nw,Aspinwall:2004jr}) Fourier-Mukai kernels for the Landau-Ginzburg monodromies $G^{-m}$, where $G$ is the generator of this monodromy (\cf Table~\ref{tab:LGDefects}). In particular for the identity defect $\wt\II^{0}$, 
the resulting kernel ($M=-d$, $k=1$) is nothing but $\OO_{\Delta X}$, the kernel of the trivial Fourier-Mukai transform.
\TABLE{
\begin{tabular}{|c|c|c|}
  \hline
  LG defect & GLSM defect & FM kernel \vphantom{\rule[-1.5ex]{0pt}{4.8ex}} \\
  \hline\hline
  $\vdots$ & $\vdots$ & $\vdots$ \\
  \hline
  $\wt I^0$ & $\widehat I^{[0,1-d,d-1,0]}$ &  $\underline{\mathcal O_{\Delta X}}$
  \vphantom{\rule[-1.5ex]{0pt}{4.8ex}} \\
  \hline
  $\wt I^{-1}$ & $\widehat I^{[d-1,0,d-1,-2]}$ &
  $\mathcal O_X \boxtimes \mathcal O_X(1)\rightarrow \underline{\mathcal O_{\Delta X}(1)}$
  \vphantom{\rule[-1.5ex]{0pt}{4.8ex}} \\
  \hline
  $\wt I^{-2}$ & $\widehat I^{[d-2,-1,d-1,-2]}$ & 
  $\Omega_{\PP^{N-1}}^1(1)\big|_X\boxtimes \mathcal O_X(1)\rightarrow
  \mathcal O_X\boxtimes \mathcal O_X(2)\rightarrow\underline{\mathcal O_{\Delta X}(2)}$
  \vphantom{\rule[-1.5ex]{0pt}{4.8ex}}\\
  \hline
  $\vdots$ & $\vdots$ & $\vdots$ \\
  \hline
  $\wt I^{-(d-1)}$ & $\widehat I^{[1,-(d-2),d-1,-2]}$ & 
  \footnotesize{${\begin{array}{l}
       \Omega^{d-2}_{\PP^{N-1}}(d-2)\big|_X\boxtimes\mathcal O_X(1) \rightarrow
          \Omega^{d-3}_{\PP^{N-1}}(d-3)\big|_X\boxtimes\mathcal O_X(2)\\
       \qquad\cdots\rightarrow\mathcal O_X\boxtimes \mathcal O_X(d-1) \rightarrow
          \underline{\mathcal O_{\Delta X}(d-1)}   
     \end{array}}$}
  \vphantom{\rule[-3.4ex]{0pt}{9ex}}\\
  \hline
  $\wt I^{-d}$ & $\widehat I^{[0,1-d,d-1,-2]}$ & $\mathcal O_{\Delta X}\{2\}$
  \vphantom{\rule[-1.5ex]{0pt}{4.8ex}} \\
  \hline
  $\vdots$ & $\vdots$ & $\vdots$ \\
  \hline
\end{tabular}
\caption{\small In this table we summarise the result of the transport of the Landau-Ginzburg monodromy defects given in the first column through the phase boundary component $N_0\times(-N_0)$ into the large volume phase. The second column contains the lifts of the defects to the gauged linear $\sigma$-model compatible with the corresponding charge window $\NNN_0\times(-\NNN_0)$. The last column shows the resulting Fourier-Mukai kernels which are obtained from the large volume defects by tensoring with $\mathcal O_X\boxtimes\mathcal O_X(d)\{1\}$. The underlined sheaves are at position zero in the complexes.}\label{tab:LGDefects}}
\subsection{Conifold and Large-Volume Monodromies in the GLSM Moduli Space}\label{sec:CandLR}
Before we start to analyse the monodromy structure in the gauged linear $\sigma$-model we should remark 
that the K\"ahler moduli space of the latter not always coincides with the K\"ahler moduli space of the corresponding non-linear $\sigma$-model. Already in its geometric phases gauged linear $\sigma$-models only capture those K\"ahler moduli of the non-linear $\sigma$-models on hyperplanes $X$, which are inherited from the ambient space\footnote{In particular for more general geometries this implies that one cannot include non-toric divisors because the gauged linear $\sigma$-model is tailor-made to describe only toric data.}.  However, if the ambient space is the projective space $\mathbb{P}^{N-1}$, and if $N>4$ then the Lefschetz hyperplane theorem guarantees that the pullback of the embedding map induces an isomorphism from the cohomology group $H^2(\PP^{N-1},\ZZ)$ of the ambient space to the cohomology group $H^2(X,\ZZ)$ of the embedding space. Therefore, for these cases we expect that 
the K\"ahler moduli spaces of gauged linear $\sigma$-model and non-linear $\sigma$-model and also monodromies in these moduli spaces agree.

Note that the above argument excludes the cubic torus in $\PP^3$ and the quartic K3 surface in $\PP^4$. In the former example, although the dimensions of the integer cohomology groups are the same, the pullback of the generator of $H^2(\PP^3,\ZZ)$ does not map to the generator of $H^2(X,\ZZ)$. As a consequence the large radius monodromy of the gauged linear $\sigma$-model is the third power of the large radius monodromy of the non-linear $\sigma$-model on the torus \cite{Jockers:2006sm}, whereas monodromies around Landau-Ginzburg and conifold point agree. In the latter example the gauged linear $\sigma$-model description only covers a slice of the non-linear $\sigma$-model K\"ahler moduli space, because for (algebraic) K3 surfaces 
$\dim H^{1,1}(X)=20$.

After this interlude about the relation between K\"ahler moduli spaces and monodromies in non-linear and gauged linear $\sigma$-models, let us now return to the discussion of the monodromies in gauged linear $\sigma$-models. By transporting the LG monodromy defects through different phase boundary components in the K\"ahler moduli space of the gauged linear $\sigma$-model, one can indeed also obtain the monodromies around conifold and large volume point in the gauged linear $\sigma$-model.

For instance, transporting the identity defect $\wt\II^0$ from the Landau-Ginzburg point into the large volume phase 
through the component $N_0\times N_d$ of the phase boundary is the same as 
first transporting it through the component $N_0\times N_{d-1}$ 
and then transporting it around the conifold point once. As discussed in Section \ref{lgmonodromylv}, 
transporting the identity defect through $N_0\times N_{d-1}$ one obtains the large volume kernel $\OO_{\Delta X}$ of the identity Fourier-Mukai transform tensored by $\OO_X\boxtimes\OO_X(-d)\{1\}$. The effect of taking it around the conifold point is then given by an action 
of the Fourier-Mukai transformation associated to the conifold monodromy. Therefore, transporting the identity defect through $N_0\times N_d$
gives rise the the Fourier-Mukai kernel of the conifold monodromy tensored by $\OO_X\boxtimes\OO_X(-d)\{1\}$.

As has been shown in \ref{glsmlift}, the lift of the identity defect $\wt\II^0$ to the GLSM compatible with $\NNN_0\times\NNN_d$, \ie the one which can be transported through the phase boundary component $N_0\times N_d$ 
is given by $\widehat\II^{[d-1,0,d,-2]}$. 
The corresponding large volume complex can be read off from \eq{kernel} to be
\beq\label{conifoldcomplex}
\left(\OO_X\boxtimes\OO_X(-d)\rightarrow\OO_{\Delta X}(-d)\right)\{1\}\,,
\eeq
giving rise to the expected Fourier-Mukai kernel
\beq\label{conifoldkernel}
\OO_X\boxtimes\OO_X\rightarrow\OO_{\Delta X}
\eeq
associated to the conifold monodromy\footnote{This is indeed also the expected monodromy in the K\"ahler moduli space of the corresponding non-linear $\sigma$-model.} (see \cite{Aspinwall:2001dz,Aspinwall:2002nw,Aspinwall:2004jr} for a target space geometric derivation of this monodromy transformation).

The Landau-Ginzburg realisation of the conifold monodromy can be obtained by transporting the large volume complex \eq{conifoldcomplex} back to the Landau-Ginzburg point through the phase boundary component $N_0\times N_{d-1}$ corresponding to the charge window $\NNN_0\times(-\NNN_0)$. This is easy to accomplish. We first note that \eq{conifoldcomplex} is a cone of two one-term complexes which we know how to transport to the LG point. Namely, $\OO_{\Delta X}(-d)\{1\}$ has been obtained by transporting $\wt\II^0$ through $N_0\times N_{d-1}$ 
to large volume in Section \ref{lgmonodromylv} above.
Moreover, $\OO_X\boxtimes\OO_X(-d)\{1\}$ is a product sheaf. As has been discussed in Section \ref{tpbranes}, the first factor $\OO_X$ can be obtained by transporting the tensor product matrix factorisation $\wt P^0$ of $W(x_i)$ to large volume through $N_0$.

Furthermore, using the fact (discussed in Appendix \ref{dualcomplexes}) that the dual $P^*$ of a matrix factorisation $P$ transported to large volume through a phase boundary component $N$ corresponding to charge window $\NNN$ gives rise to the dual of the large volume complex obtained by transporting $P$ to large volume through the component $-N$ associated to the dual charge window $-\NNN$ tensored by $\OO_X(-d)\{1\}$, we see that 
$\OO_X(-d)\{1\}\cong\OO_X^*(-d)\{1\}$ can be obtained by transporting $(\wt P^0)^*$ to large volume through phase boundary component $N_{d-1}$. Thus, the realisation of the conifold monodromy is given by the cone
\beq\label{mfconicone}
{\rm Cone}\left(\wt r:\wt P^0(x)\otimes \left(\wt P^0(y)\right)^*\rightarrow\wt\II^0\right)\,,
\eeq
where $\wt P^0(x)$ denotes the respective equivariant tensor product matrix factorisation of $W(x_1,\ldots,x_N)$,
and 
the map $\wt r$ is induced by the identity map $\id:P^0\rightarrow P^0$ by means of the folding isomorphism \eq{foldingformula}. In physics terminology this is the tachyon condensation of the sum of defect matrix factorisations 
\beq
\wt P^0(x)\otimes \left(\wt P^0(y)\right)^*\{1\}\oplus\wt\II^0
\eeq
with tachyon associated to $\wt r$. In Section \ref{conifoldsection} we will discuss the conifold monodromies in more detail and more generality.

To obtain the inverse of the large volume monodromy one has to compose the generator of the conifold monodromy with the inverse of the generator $G$ of the Landau-Ginzburg monodromy. The corresponding large volume complex can be obtained by transporting the defect $\wt\II^1$ through the 
phase boundary component $N_0\times N_d$. The lift of $\wt\II^1$ compatible with the corresponding charge window 
$\NNN_0\times(-\NNN_0+1)$ is given by $\widehat\II^{[0,1-d,d,0]}$.
The large volume complex can then be read off from \eq{kernel} to be
\beq
\OO_{\Delta X}(-1-d)\{1\}\,.
\eeq
Tensoring by $\OO_X\boxtimes\OO_X(d)\{-1\}$ we obtain the Fourier-Mukai kernel $\OO_{\Delta X}(-1)$ of the inverse of the large-volume monodromy. This is also the expected result for the monodromy around the large volume limit in the K\"ahler moduli space of the gauged linear $\sigma$-model \cite{Horja,Aspinwall:2001dz,Aspinwall:2002nw}. Note that, as alluded to above, in the case that $X$ is the cubic torus in $\PP^2$, \ie $N=d=3$, this monodromy does not coincide with the large volume monodromy in the K\"ahler moduli space of the corresponding non-linear $\sigma$-model, but is its third power.
\section{Conifold Monodromy}\label{conifoldsection}
Let us now discuss the conifold monodromy in more detail.
A conifold point in the K\"ahler moduli space of a non-linear $\sigma$-model on a Calabi-Yau manifold $X$, is characterised by the property that at this point a certain B-type BPS brane $Q$ becomes massless.  This means that its (quantum) world-volume in the Calabi-Yau space vanishes while the (quantum) world-volumes of other B-type branes remain finite \cite{Greene:1996tx}. 
Such points are easily identified in the mirror geometry, where the vanishing quantum cycle of the B-brane $Q$ arises classically as a singularity in the mirror Calabi-Yau manifold. Locally, the latter can be described by the 
geometry of a singular conifold \cite{Candelas:1989js}.

B-type D-branes transform under 
the 
monodromy around conifold points in K\"ahler moduli space in a non-trivial way. The monodromy action on 
D-brane charges is encoded in the periods in the vicinity of the conifold point \cite{Kontsevich:1994,Greene:1996tx,Brunner:2001eg}. But the analysis can be extended beyond the level of D-brane charges to the category of topological B-branes. It turns out that the conifold monodromy acts on the large volume realisation $D^b(X)$ of this category by means of the Fourier-Mukai transformation $\Phi^{{\cal K}_{\cal Q}^{\rm C}}$ with kernel
\begin{equation} \label{eq:ConiKernel}
     {\cal K}_{\cal Q}^{\rm C} = 
     {\rm Cone}(r:{\cal Q} \boxtimes {\cal Q}^{\lor} \rightarrow {\cal O}_{\Delta X}) \ ,
\end{equation}
determined by the large volume complex ${\cal Q}$ associated to the B-brane $Q$ \cite{Aspinwall:2001dz,Seidel:2000}. Here ${\cal Q}^\lor$ denotes the dual of ${\cal Q}$, and ${\cal O}_{\Delta X}$ is structure sheaf of the diagonal $\Delta X\subset X\times X$. 
The map $r$ is the restriction map to the diagonal $\Delta X$. If for instance ${\cal Q}=\OO_X$, then the map $r$ restricts $\OO_{X\times X}=\OO_X\boxtimes\OO_X$ to $\OO_{\Delta X}$.

In general, the map $r$ which is an element of ${\rm Hom}({\cal Q}\boxtimes{\cal Q}^{\lor},{\cal O}_{\Delta X})$, is induced from the identity map $\id\in{\rm Hom}({\cal Q},{\cal Q})$ by means of the following chain of isomorphisms: The complex ${\cal Q}$ is quasi-isomorphic to its image $\Phi^{\OO_{\Delta X}}({\cal Q})$ under the trivial Fourier-Mukai transform. Therefore ${\rm Hom}({\cal Q},{\cal Q})$ is isomorphic to ${\rm Hom}({\cal Q},\Phi^{{\cal O}_{\Delta X}}({\cal Q}))$, which in turn is isomorphic to ${\rm Hom}({\cal Q}\boxtimes{\cal Q}^{\lor},{\cal O}_{\Delta X})$ (\cf equation \eq{FMHomformula}).

The action of the Fourier-Mukai transformation $\Phi^{{\cal K}_{\cal Q}^{\rm C}}$ associated to the kernel \eq{eq:ConiKernel} on a complex ${\cal E}$ can also be represented in the following way \cite{Seidel:2000,Jockers:2006sm}
\begin{equation}\label{coniaction}
   {\cal E}\mapsto {\rm Cone}\left({\rm ev}: {\rm Hom}({\cal Q},{\cal E})\otimes{\cal Q}\rightarrow {\cal E} \right) \ ,
\end{equation} 
where `$\,{\rm ev}\,$' denotes the evaluation map. This expression also makes more transparent that the advocated Fourier-Mukai transformation $\Phi^{{\cal K}_{\cal Q}^{\rm C}}$ indeed reduces to the expected map on the level of D-brane charges.

In this paper, we focused our considerations on non-linear $\sigma$-models whose target spaces $X$ 
are projective Calabi-Yau hypersurfaces. The K\"ahler moduli spaces of the gauged linear $\sigma$-model realisation of these models exhibit a conifold singularity, at which the (quantum) world-volume of the entire target space is zero.

The monodromy around this point was derived in Section~\ref{sec:CandLR} by lifting the identity defect to the gauged linear $\sigma$-model and transporting it around the singularity. In the large volume phase we indeed obtained the Fourier-Mukai transformation with kernel \eq{conifoldkernel} which is of the general form ${\cal K}_{\cal Q}^{\rm C}$ where ${\cal Q}=\OO_X$ is the large volume complex representing the B-brane which becomes massless at the conifold point. 

Thus, this approach provides an independent derivation of the conifold monodromy for this particular case. Moreover it also yields a realisation of this monodromy at the Landau-Ginzburg point. Namely, we obtained the cone \eq{mfconicone} as the matrix factorisation representing the defect which realises this monodromy. The structure of this matrix factorisation is very reminiscent of the structure of the Fourier-Mukai kernel ${\cal K}_{\cal Q}^{\rm C}$.  In the following, we will generalise these Landau-Ginzburg defects to arbitrary B-branes $Q$ and show that transported to large volume they reproduce the Fourier-Mukai kernels ${\cal K}_{\cal Q}^{\rm C}$. Hence, we obtain a Landau-Ginzburg realisation of arbitrary conifold monodromies.
\subsection{Conifold-like Defects in general CFTs}
As preparation,  we would like to give a non-technical description of
a class of defects arising in any $N=(2,2)$ superconformal field theory, whose action on 
boundary conditions mimics conifold monodromy transformations. The conifold monodromy defects which we will present in the next section are a concrete example of this general construction in the context of Landau-Ginzburg models.

As described in the introduction, conifold monodromies can be understood as follows. 
Any number of D-branes $Q$ which become massless at a conifold point can be created there at no cost in energy. When one transports a massive probe D-brane $E$ around this point, it forms bound states with these D-branes, provided there are suitable open string tachyons between $Q$ and $E$. This is exactly what is described by formula \eq{coniaction} in the context of large volume non-linear $\sigma$-models. 

Therefore, a defect mimicking such a monodromy transformation has to map a boundary condition $E$
to itself, and in addition it has to create a copy of a chosen boundary condition $Q$
for every possible tachyon between $Q$ and $E$. Moreover, it has to form a bound state
of this collection of boundary conditions.

Defects which upon fusion with a boundary condition $E$ 
create the above constituent boundary conditions before bound state formation
are very easy to find.
First of all, the identity defect ${\rm Id}$, which exists in all CFTs maps any boundary condition $E$ to itself. The second constituent can be obtained by fusion with a totally reflective defect. 

Totally reflective defects are defects which do not allow the transmission of excitation
from one of the adjacent CFTs to the other. Hence, they are defined by boundary conditions in each of these CFTs. For instance, one can choose the boundary condition $Q$ on one side of the defect and the world-sheet parity dual $Q^*$ of $Q$ on the other side. If the corresponding defect $T_Q=Q\otimes Q^*$ is fused
with the boundary condition $E$, it creates a copy of the boundary condition
$Q$ for every tachyon between $Q$ and $E$, or to be more precise, it turns $E$ into
the boundary condition $\HH^*(Q,E)\otimes Q$. Thus, identity defect ${\rm Id}$
and the totally reflective defect $T_Q$ indeed provide the building blocks needed to produce conifold-like transformation on boundary conditions, \cf \eq{coniaction}.
\FIGURE{
\includegraphics[height=4.5cm]{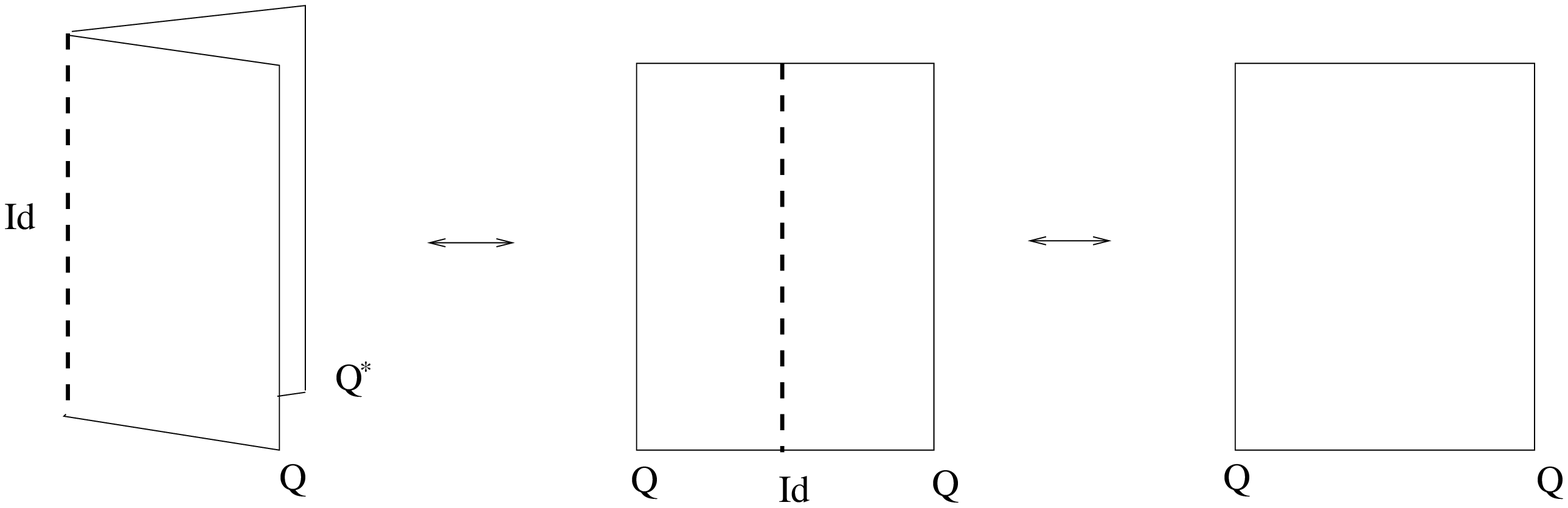}
\caption{\small The spectrum of boundary fields of a boundary condition $Q$ (right)
always contains the identity. This yields a universal element in the spectrum of defect-changing fields between the totally reflective defect $T_Q$ and the identity defect (left).}
\label{fig:idtensor}}

The bound state formation of this collection of boundary conditions is then induced by a 
bound state formation of the anti-defect $T_Q\{1\}$ whose R-charges have been shifted by $1$ relative to $T_Q$ and the trivial defect ${\rm Id}$.

Indeed, there is a universal defect-changing field in the spectrum between these two defects which 
can be used to trigger the corresponding tachyon condensation. It can be described by means of the folding trick, \cf Figure~\ref{fig:idtensor}. Being a totally reflective defect, one can cut open the world-sheet along the defect $T_Q$. In particular, if ${\rm Id}$ and $T_Q$ run parallel along a cylinder, the latter can be unfolded to a strip with boundary conditions $Q$ at both its edges.
Since ${\rm Id}$ is the trivial defect, the spectrum of defect-changing fields between ${\rm Id}$ and $T_Q$ is therefore isomorphic to the spectrum of boundary fields on the boundary condition $Q$.  (This is a special case of the folding formula \eq{foldingformula}.)

Thus, the identity boundary field on $Q$ gives rise to a universal bosonic defect-changing field between $T_Q$ and ${\rm Id}$, which is promoted to a fermionic field between the shifted defect $T_Q\{1\}$ and ${\rm Id}$.
This defect-changing field can be used to form a bound state $C_Q$ of the defects $T_Q\{1\}$ and ${\rm Id}$, and upon fusion of $C_Q$ with a boundary condition $E$ it induces exactly the tachyons between $E$ and the collection of boundary conditions $Q$ created by $T_Q$ (\ie $\HH^*(Q,E)\otimes Q$), which trigger the desired bound state formation.
It is the exact analogue of the map ${\rm ev}$ in \eq{coniaction}.

Hence, the bound state defect $C_Q$ realises conifold-like actions on boundary conditions in any $N=(2,2)$ superconformal field theory.

To relate this general construction to the derivation of the special conifold defects in Section \ref{sec:CandLR}, note that the conifold defects
\eqref{mfconicone} obtained there in the context of Landau-Ginzburg orbifolds indeed are of the form $C_Q$, where $Q$ is given by the matrix factorisation $\wt P^0$. The map $\wt r$ is exactly the universal defect-changing field discussed above.
\subsection{Conifold Defects in the Landau-Ginzburg Phase}
After this digression into the context of general $N=(2,2)$ superconformal field theories, we will turn back to conifold monodromies in non-linear $\sigma$-models. 
In the following, we will realise the Fourier-Mukai transformations $\Phi^{{\cal K}_{\cal Q}^{\rm C}}$ representing monodromies around conifold points associated to B-type D-branes with arbitrary large volume complex ${\cal Q}$ 
as defects in the Landau-Ginzburg phase. 

Guided by the above discussion, we first need to identify the equivariant matrix factorisation
\begin{equation}
     Q: \xymatrix{{Q}_1 \ar@<0.5ex>[r]^{q_1} & 
                                                          {Q}_0 \ar@<0.5ex>[l]^{q_0}  } \ ,     
\end{equation}
of $W(x_1,\ldots,x_N)$ 
which in the Landau-Ginzburg phase represents the BPS brane that becomes massless at the chosen conifold locus. 
Let $Q^*$ be the dual of the matrix factorisation $Q$ defined as in \eq{eq:MFdual}. Note that the representations of
the orbifold group $\Gamma$ and $R$-symmetry group on the dual matrix factorisation $Q^*$ are given by the dual of the representations on $Q$. 

Now let us denote by $Q(y)$ the matrix factorisation $W(y_1,\ldots,y_N)$ obtained from $Q$ by replacing the variables $x_i$ by $y_i$, and set $Q(x)=Q$. The tensor product (as defined in \eq{eq:MFtensor}) $Q(x)\otimes Q^*(y)$ of the matrix factorisations $Q(x)$ and the dual of $Q(y)$ yields a matrix factorisation of $W(x_1,\ldots,x_N)-W(y_1,\ldots,y_N)$ which will eventually corresponds to the large volume complex ${\cal Q}\boxtimes{\cal Q}^\lor$ in the conifold Fourier-Mukai kernel ${\cal K}_{\cal Q}^{\rm C}$ defined in \eq{eq:ConiKernel}.
We denote this tensor product matrix factorisation by $T_Q$. Concretely, it reads
\begin{equation}\label{eq:TQ}
      {T}_{Q}={Q}(x) \otimes{Q}^{*}(y):
         \xymatrix{
           {\begin{matrix} Q_1(x) \otimes  Q_0^*(y)\\ \oplus\\ Q_0(x) \otimes Q_1^*(y) \end{matrix}}  
          \ar@<0.5ex>[rrrrr]^{t_1=
             \left(\begin{smallmatrix}
                \phantom{-}q_1(x) \otimes {\rm id} & \phantom{-}{\rm id} \otimes q_0^*(y) \\
                     \phantom{-}{\rm id} \otimes q_1^*(y) & \phantom{-}q_0(x)\otimes {\rm id}
             \end{smallmatrix}\right)}
           &&&&&
          \ar@<0.5ex>[lllll]^{t_0=
                 \left(\begin{smallmatrix}
                    \phantom{-}q_0(x)\otimes{\rm id} & -{\rm id}\otimes q_0^*(y) \\
                    -{\rm id}\otimes q_1^*(y) & \phantom{-}q_1(x)\otimes {\rm id} 
                 \end{smallmatrix}\right)}
           {\begin{matrix} Q_0(x) \otimes Q_0^*(y)\\\oplus\\ Q_1(x) \otimes Q_1^*(y) \end{matrix}}  }
\end{equation}
This is a $\Gamma\times\Gamma\times{\rm U}(1)_R$-equivariant matrix factorisation with representations induced by the ones on the factors $Q(x)$ and $Q^*(y)$.

The next ingredient in the defect \eq{mfconicone} realising the monodromy around the conifold point associated to $\OO_X$ is the matrix factorisation $\wt\II^0$ representing the identity defect. As has been discussed in Section~\ref{lglgmon} this corresponds to the trivial Fourier-Mukai transform at large volume, which has the diagonal structure sheaf $\OO_{\Delta X}$ as kernel. 

Finally, the map $\wt r$ in \eq{mfconicone} is replaced by the analogous map $\wt r\in\HH^0_{\rm orb}(T_Q,\wt\II^0)$ represented by the maps $\wt r_s$ in the commutative diagram
\begin{equation}
     \xymatrix{  (T_Q)_1 \ar@<0.5ex>[r]^{t_1}  \ar[d]_{\tilde r_1} & \ar@<0.5ex>[l]^{t_0} \ar[d]^{\tilde r_0} (T_Q)_0 \\
                       \wt I_1 \ar@<0.5ex>[r]^{\wt\imath_1} & \wt I_0 \ar@<0.5ex>[l]^{\wt\imath_0} } 
\end{equation}
which is induced by the identity morphism of the matrix factorisation $Q$ by means of the folding isomorphisms
$\HH^0(Q(x),Q(x))\cong\HH^0(Q(x),\wt\II^0* Q(y))\cong\HH^0(T_Q,\wt\II^0)$, \cf \eq{foldingformula}. This map corresponds to the restriction map $r$ in the Fourier-Mukai kernel \eq{eq:ConiKernel} at large volume.
Its explicit computation is given in Appendix~\ref{sec:tachyon}.

Now we have all the ingredients at hand to state the matrix factorisation representing the the conifold defect at the Landau-Ginzburg point. As in \eq{mfconicone} it is given by the cone construction with respect to the map $\wt r$:
\begin{equation} \label{eq:MFConiCone}
     C_Q \,=\,
     {\rm Cone}\left(\wt r:\  Q(x)\otimes Q^*(y) \rightarrow {\wt I}^0\right) \ .
\end{equation}
In physics language this construction corresponds to the topological condensate \cite{Sen:1998sm} of the tensor product defect
${T}_{Q}\{1\}$ and $\wt{I}^0$, triggered by the fermionic open-string cohomology element corresponding to $\wt{r}$ . Note that here ${T}_Q\{1\}$ denotes the anti-defect of ${T}_{Q}$ whose $R$-charges have been shifted by $1$. By this spectral flow, the bosonic morphism $\wt r$ has become fermionic \cite{Witten:1998cd}. The final result of this condensation process yields the matrix factorisation ${C}_{Q}$ for the conifold defect, which more concretely reads
\begin{equation}
      {C}_{Q}:
         \xymatrix{
           {\begin{matrix} \wt I_1\\\oplus\\ \wt T_0 \end{matrix}}  
          \ar@<0.5ex>[rrr]^{c_1=
             \left(\begin{smallmatrix}
                \wt\imath_1 & \phantom{-}\tilde r_0 \\
                0 & -t_0
             \end{smallmatrix}\right)}
           &&&
          \ar@<0.5ex>[lll]^{c_0=
                 \left(\begin{smallmatrix}
                   \wt\imath_0 & \phantom{-}\tilde r_1 \\
                    0 & -t_1 
                 \end{smallmatrix}\right)}
           {\begin{matrix} \wt I_0\\ \oplus\\\wt T_1 \end{matrix}}  }
\end{equation}
Note that due to the constant entries in the topological tachyon $\wt r$, 
this matrix factorisation can always be reduced to smaller rank 
by removing trivial brane-anti-brane pairs \cite{Govindarajan:2005im}.

In this way we obtain candidates for the Landau-Ginzburg defect realisations of general conifold monodromies. In the next section we will show that they indeed give rise to the desired Fourier-Mukai transformations $\Phi^{{\cal K}_{\cal Q}^{\rm C}}$ with kernels \eq{eq:ConiKernel} when transported into the large volume phase. 
\subsection{Transport to the Large Volume Phase}
In order to transport the conifold defect \eq{eq:MFConiCone} to the large volume phase we again follow the steps used in Section \ref{sec:CandLR}. We first observe that if two matrix factorisations $P$ of $W(x_i)$ and $Q$ of $W(y_i)$ give rise to large volume complexes ${\cal P}$ and ${\cal Q}$ when transported into the large volume phase through phase boundaries $N$ and $N\p$ respectively, the tensor product $P\otimes Q$ gives rise to the exterior product ${\cal P}\boxtimes{\cal Q}$ when transported to large volume through the phase boundary segment $N\times N\p$.
This can directly be seen by comparing the unbounded complex obtained from the lift of the tensor product $P\otimes Q$ to the exterior product of the two unbounded complexes ${\cal P}$ and ${\cal Q}$. This general relation can be formulated as
\begin{equation} \label{eq:ExtTensor}
     {\rm LV}_N(P)\boxtimes{\rm LV}_{N\p}(Q)\cong{\rm LV}_{N\times N\p}(P\otimes Q)\,.
\end{equation}
Furthermore, we use the relation \eq{eq:dualrel}
\beq
{\rm LV}_{-N}(P^*)\cong\left({\rm LV}_N(P)\right)^\lor\otimes\OO_X(d)\{-1\}
\eeq
between duality of matrix factorisations at the Landau-Ginzburg point 
and duality of complexes of coherent sheaves at large volume derived in Appendix \ref{dualcomplexes}. Here
$-N$ is the phase boundary component whose associated charge window 
is the dual $-\NNN$ of the one $\NNN$ associated to $N$.

Using these two relations, one easily sees that if the matrix factorisation $Q$ corresponds to the large volume complex ${\cal Q}$ when transported from the Landau-Ginzburg to the large volume phase through the phase boundary component $N$, $T_Q$ gives rise to 
\beq
{\rm LV}_{N\times N^*}\cong {\cal Q}\boxtimes\left({\cal Q}^\lor\otimes\OO_X(-d)\{1\}\right)
\eeq
upon parallel transport through the phase boundary segment $N\times N^*$. 

Using that the identity defect transports to the large volume complex
\beq
LV_{N\times N^*}(\wt\II^0)\cong\OO_{\Delta X}(-d)\{1\}
\eeq
which was shown in Section \ref{lgmonodromylv}, we finally arrive at the large volume complex
\beq\label{LVconikernel}
{\rm LV}_{N\times N^*}(C_Q)\cong{\rm Cone}\left(r:{\cal Q}\boxtimes\left({\cal Q}^\lor\otimes\OO_X(-d)\{1\}\right)
\rightarrow \OO_{\Delta X}(-d)\{1\}\right)
\eeq
arising by transporting the defect $C_Q$ of \eq{eq:MFConiCone} through the phase boundary segment $N\times N^*$ to the large volume phase.  Here, the restriction map $r$ arises in the following way: In the Landau-Ginzburg phase the corresponding morphism $\wt r$ of matrix factorisations
is induced from the identity map in ${\rm Hom}(Q,Q)$. As a result the large volume counterpart must 
be induced from the identity morphism ${\rm Hom}({\cal Q},{\cal Q})$ as well. This, however, is precisely the definition of the restriction map $r$ in eq.~\eqref{eq:ConiKernel}. 

Tensoring the large volume kernel \eq{LVconikernel} by ${\cal O}_X\boxtimes{\cal O}_X(d)\{-1\}$ to obtain the associated Fourier-Mukai kernel (\cf \eq{eq:KerDefRelation}), we indeed find that the defect $C_Q$ at the Landau-Ginzburg point 
realises the Fourier-Mukai transformation
$\Phi^{{\cal K}_{\cal Q}^{\rm C}}$ on the category of the large volume B-branes which is associated to the conifold kernel ${\cal K}_{\cal Q}^{\rm C}$. This shows that $C_Q$ indeed represents the monodromy around the conifold point at which the D-brane associated to $Q$ becomes massless.
\section{Discussion}\label{discussionsection}
In this paper we have discussed B-brane monodromy transformations from the world-sheet point of view.
Our construction of monodromies can be understood as an application of the idea put forward in \cite{Brunner:2007ur} that the effect which bulk perturbations of conformal field theories have on boundary conditions imposed on the world-sheet boundaries
can be realised by fusion with specific defects
associated to the perturbations.  

We have constructed the defects corresponding to deformations along closed loops in K\"ahler moduli space of non-linear $\sigma$-models, which start and end at Landau-Ginzburg points. At these points the study of B-type defects and D-branes is particularly simple, because they can be elegantly described by means of matrix factorisations.

Using the description of B-type D-branes in gauged linear $\sigma$-models developed in \cite{Herbst:2008jq} we have shown that the action of any B-type defect at the Landau-Ginzburg point is realised by a Fourier-Mukai transformation on the B-brane category at large volume. Moreover, the particular Fourier-Mukai transformations arising in this way from our monodromy defects indeed agree with the ones obtained from a more target space geometric approach to monodromies in \cite{Horja,Aspinwall:2001dz,Aspinwall:2002nw}. 

This gives further support to the idea that the effect of bulk perturbations on boundary conditions can be efficiently described using defects. A different type of examples, namely relevant flows between $N=2$ minimal models has been analysed in \cite{Brunner:2007ur}.

We have focused our considerations on non-linear $\sigma$-models on Calabi-Yau hypersurfaces in projective space, and also on the part of their K\"ahler moduli spaces which can be realised by means of a gauged linear $\sigma$-model with ${\rm U}(1)$ gauge group. 
It would be interesting to adapt our constructions to more general situations such as models with a more complicated phase structure. In particular, the lift to the gauged linear $\sigma$-model allows to derive all defects arising from monodromies around singularities in K\"ahler moduli space which are captured by the GLSM, by transporting the identity defect around the respective loops in K\"ahler moduli space. In this way, also more complicated monodromy groups could be analysed.

Our finding that fusion of defects at Landau-Ginzburg points is
realised by means of Fourier-Mukai transformations in the large volume
regime can also be turned around to provide a complementary
perspective on functors on B-brane categories. Fusion of defects in
Landau-Ginzburg models seems to be conceptually simpler and in some
cases easier to work with in practice than 
the corresponding Fourier-Mukai transformations at large volume. 
As the example of the ``quantum symmetry defects''
discussed in Section \ref{lglgmon} strikingly demonstrates, 
there are functors which seem to be rather complicated at large volume but have a very simple defect realisation at the Landau-Ginzburg point. 

In particular, the Landau-Ginzburg realisation 
provides a new and potentially simpler method to calculate D-brane monodromies
\cite{Aspinwall:2001dz,Distler:2002ym,Jockers:2006sm}.
Interesting cases to study include the action of monodromies on D0-branes supported
at specific points of Calabi-Yau target spaces, as well as examples exhibiting K-theory torsion.
%
%
%
%
\subsection*{Acknowledgements}
H.~J.~ and D.~R.~ would like to thank the Banff International Research Station and the organisers of the workshop on matrix factorisations for a stimulating conference. I.~B.~ thanks the NHETC at Rutgers University
for hospitality.
We would also like to thank E.~Diaconescu, M.~Herbst and K.~Hori for useful discussions. The work of                    
I.~B.~ is supported by a EURYI award of the European science foundation and       
furthermore by the Marie-Curie network Forces Universe                             
(MRTN-CT-2004-005104). H.~J.~ is supported by the Stanford Institute for Theoretical              
Physics.
D.~R.~ is supported by a DFG research fellowship and partially by DOE-grant DE-FG02-96ER40959.
\appendix
\section{Dual Matrix Factorisations and Dual Complexes}\label{dualcomplexes}
Lifting a matrix factorisation $P$ and its dual $P^*$ to the GLSM and transporting it into the large volume phase through the phase boundary component $N$ with associated charge window $\NNN$ and its dual $N^*$ with charge window $-\NNN$ respectively, one obtains two complexes ${\cal P}:={\rm LV}_N(P)$ and $line{\cal P}:={\rm LV}_{N^*}(P^*)$. In this appendix we will show that they satisfy the following relations
\begin{equation} \label{eq:dualrel}
{\rm LV}_N(P)\cong\left({\rm LV}_{N^*}(P^*)\right)^\lor\otimes\OO_X(-d)\{1\}
     \ , \quad
\left({\rm LV}_N(P)\right)^\lor\cong{\rm LV}_{N^*}(P^*)^\lor\otimes\OO_X(d)\{-1\}\,.
\end{equation}
Here `\,$\cong$\,' means quasi-isomorphic, and `\,${}^\lor$\,' refers to the dual complex of coherent sheaves. 

The dual $P^*$ of the matrix factorisation $P$ is defined in equation~\eqref{eq:MFdual}, and it carries the
duals $\rho^*$, $(\rho^R)^*$ of the representations $\rho$ and $\rho^R$ of the orbifold group $\Gamma=\ZZ_d$ and the $R$-symmetry group ${\rm U}(1)_R$ defined on $P$:
\begin{equation} \label{eq:dualrep}
    (\rho^*)_s(\gamma) = (\rho_s)^*(\gamma^{-1}) \ , \qquad ((\rho^R)^*)_s(\varphi) = (\rho^R)^*_s(-\varphi) \ .
\end{equation}
As reviewed in Section~\ref{glsm} the large radius complex, ${\cal P}={\rm LV}_N(P)$ obtained by transporting the Landau-Ginzburg B-brane represented by the matrix factorisation $P$ through the phase boundary component $N$ into the large volume phase has the general form
\begin{equation} \label{eq:ComplexEP}
      {\cal P}\,: \ 
      \xymatrix{ \cdots \ar[r] & 
      {\cal P}^{k-1} \ar[r]^{\wp_{k-1}} &
      {\cal P}^k \ar[r]^{\wp_k} &
      {\cal P}^{k+1} \ar[r]^{\wp_{k+1}} & \cdots } \ .
\end{equation}
The sheaves ${\cal P}^k$ which appear in the complex at grading $k$ are given by
\begin{equation}
     {\cal P}^k = \bigoplus_{i \in A_k} {\cal O}_X\left(\tfrac{d}{2}(k-r_i)+q_i\right) \ ,
\end{equation}
in terms of the GLSM $R$-charges $r_i$ defined in Section~\ref{sec:GLSMtoLG} and the index sets
\begin{equation}
     A_k= 
     \left\{\, i \,\vphantom{\frac12}\right|\left.\ 0 \le \frac12\left(k-r_i\right)\in\ZZ \right\} \ .
\end{equation}
The ${\rm U}(1)_R$-charges $q_i$ lie in the charge window ${\cal N}$ associated to the phase boundary component $N$.
Finally, the maps $\wp_k$ are truncations of the maps $p_s$ defining the matrix factorisations $P$.

The index set $A_k$ arises from the fact that in order to go to the
large volume phase
we need to integrate out the field $\mathfrak{p}$. On the boundary
however, the field 
$\mathfrak{p}$ gives rise to a boundary interaction, which generates
a Fock space for each index~$i$ \cite{Herbst:2008jq}. The Fock space vacua have $U(1)\times
U(1)_R$ charges
$(q_i, r_i)$, whereas their $j$-th excited states
have $U(1)\times U(1)_R$ 
charges $(q_i+j d, r_i +2 j)$. The index set
$A_k$ is a bookkeeping
device for all the states in the Fock spaces. At large volume these states give rise to the 
line bundles $\OO_X(q_i+jd)$ in the complex representing the D-brane under consideration.

There are a few comments in order. First, we observe that the complex~\eqref{eq:ComplexEP} is bounded to the left because for sufficiently small $k$ the index sets $A_k$ are all empty. On the other hand, for $k$ sufficiently large 
the complex becomes two-periodic in the sense that  $\wp_k=p_{k\,{\rm mod}\,2}$. 

Analogously, lifting dual matrix factorisation $P^*$ to the GLSM and transporting it  through the phase boundary component $N^*$ corresponding to the dual charge window  $\NNN^*=-{\cal N}$, the resulting complex reads
 \begin{equation} \label{eq:ComplexEPStar}
      \overline{\cal P}\,: \ 
      \xymatrix{ \cdots \ar[r] & 
      \overline{\cal P}^{k-1} \ar[r]^{\overline{\wp}_{k-1}} & 
      \overline{\cal P}^k \ar[r]^{\overline{\wp}_{k}} &
      \overline{\cal P}^{k+1} \ar[r]^{\overline{\wp}_{k+1}} & \cdots } \ ,
\end{equation}
with
\begin{equation}
     \overline{\cal P}^k = \bigoplus_{i \in A^*_k} {\cal O}_X\left(\tfrac{d}{2}(k+r_i)-q_i\right) \ , \qquad
     A^*_k= \left\{\, i \,\vphantom{\frac12}\right|\left.\ 0 \le \frac12\left(k+r_i\right)\in\ZZ \right\} \ .
\end{equation}
Here the signs of the ${\rm U}(1)$- and ${\rm U}(1)_R$-charges are reversed relative to the ones appearing in the definition of ${\cal P}$, because $P^*$ carries the dual representations of $\Gamma=\ZZ_d$ and ${\rm U}(1)_R$,
and $P^*$ was lifted in a way compatible with the dual charge window $-\NNN$.
The maps $\overline{\wp}_k$ in 
this complex are now truncations of the maps
$(p^*)_0$ and $-(p^*)_1$ defining the dual $P^*$ of the matrix factorisation $P$. 

In order to determine the relation between the complexes 
\eqref{eq:ComplexEPStar} and \eqref{eq:ComplexEP} 
we form the dual complex, $\overline{\cal P}^\lor$:
 \begin{equation} \label{eq:ComplexEPStarLor}
      \overline{\cal P}^\lor\,: \ 
      \xymatrix{ 
         \cdots \ar[r] & 
         (\overline{\cal P}^\lor)^{-(k+1)} \ar[r]^{\overline{\wp}_k^*} &
         (\overline{\cal P}^\lor)^{-k} \ar[r]^{\overline{\wp}_{k-1}^*}  &
         (\overline{\cal P}^\lor)^{-(k-1)} \ar[r] & \cdots } \ .
\end{equation}
The maps $\overline{\wp}_k^*$ in the dual complex are induced by the maps $(p_s^*)^*=p_s$. 
The coherent sheaves in the dual complex at grading $-k$ are the dual sheaves of the original complex at grading $k$. Therefore, for the coherent sheaves in the dual complex at grading $k$ we arrive at
\begin{equation}
     (\overline{\cal P}^\lor)^k = \bigoplus_{i \in A^{*\,\lor}_k} {\cal O}_X\left(\tfrac{d}{2}(k-r_i)+q_i\right) \ , \qquad
     A^{*\,\lor}_k= \left\{\, i \,\vphantom{\frac12}\right|\left.\ 0 \ge \frac12\left(k-r_i\right)\in\ZZ \right\} \ .
\end{equation}
By construction this dual complex is bounded to the right and two-periodic to the left with $(\overline{\wp}^*)_k=p_{k\,{\rm mod}\,2}$ for $k$ sufficiently small. Note that the left two-periodic part of $\overline{P}^\lor$ is given by the maps $p_s$, as is the 
right two-periodic part of ${\cal P}$. 
Thus, it is tempting to combine these two complexes to form a single two-periodic complex unbounded to both sides. This is not quite possible however, because the index sets $A_k$ and $A_k^{*\,\lor}$ are not complementary. But this can be easily remedied
by tensoring the complex $\overline{\cal P}^\lor$
with the one-term complex ${\cal O}_X(-d)\{2\}$. The resulting complex
 \begin{equation} \label{eq:ComplexEPStarLorST}
      {\cal F}:=\overline{\cal P}^\lor{\otimes}{\cal O}_X(-d)\{2\} \,: \ 
      \xymatrix{ \cdots \ar[r] & {\cal F}^{k-1} \ar[r] & {\cal F}^k \ar[r] & {\cal F}^{k+1} \ar[r] & \cdots } \ ,
\end{equation}
now consists of sheaves
\begin{equation}
     {\cal F}^k = \bigoplus_{i \in B_k} {\cal O}_X\left(\tfrac{d}{2}(k-r_i)+q_i\right) \ , \qquad
     B_k= \left\{\, i \,\vphantom{\frac12}\right|\left.\ 0 > \frac12\left(k-r_i\right)\in\ZZ \right\} 
\end{equation}
defined by index sets $B_k$ which are complementary to $A_k$.

Let us now analyse the consequences of this observation. As indicated we can now combine the two complexes ${\cal P}$ and $\overline{\cal P}^\lor{\otimes}{\cal O}_X(-d)\{2\}$ to form the announced unbounded two-periodic complex, which we denote by ${\cal Z}$. This is achieved in two steps. We first take the direct sum of these two complexes and then supplement the maps, which we collectively denote by $\delta\wp$ in the resulting complex at various gradings in order to obtain the desired two-periodicity. The described procedure, however, is simply the cone construction, \ie the resulting complex ${\cal Z}$ is formally given by
\begin{equation} \label{eq:coneZ}
     {\cal Z}= {\rm Cone}\left(
     \delta \wp : \ 
     \overline{\cal P}^\lor{\otimes}{\cal O}_X(-d)\{1\} 
     \rightarrow
     {\cal P} \right) \ .
\end{equation}
But this complex is nothing but the unbounded two-periodic complex defined by the matrix factorisation $P$, 
with maps $z_k=p_{k\,{\rm mod}\,2}$. 
Therefore, for all $k$ the image of the map $z_k$ equals the kernel of the map $z_{k+1}$, and hence the complex ${\cal Z}$ has no cohomology and is quasi-isomorphic to the null complex.
This in turn implies that the constituents $\overline{\cal P}^\lor{\otimes}{\cal O}_X(-d)\{1\}$ and ${\cal P}$ are quasi-isomorphic to each other, which implies the advocated relations~\eqref{eq:dualrel}.
\section{Restriction Map Tachyon} \label{sec:tachyon}
Here we derive the bosonic  morphism $\wt{r}$ between the tensor product matrix factorisations $T_{\wt Q}$ defined 
in \eq{eq:TQ} and the matrix factorisation $\wt\II^0$ representing the identity defect. It is an element of the BRST-cohomology group ${\cal H}^0_{\rm orb}(T_{\wt Q},\wt{I}^0)$ which is induced from the identity map in ${\cal H}^0_{\rm orb}(\wt Q,\wt Q)$ by means of the folding isomorphism \eq{foldingformula}. First, we construct the cohomology element $r=(r_0,r_1)$ of the corresponding matrix factorisations in the unorbifolded theory. It arises in the following commutative diagram:
\begin{equation} \label{eq:tachyon}
     \xymatrix{
        {\begin{matrix}Q_1\otimes Q_0^*\\\oplus\\Q_0\otimes Q_1^*\end{matrix}} 
        \ar@<0.5ex>[rrrrr]^{t_1
            =\left(\begin{smallmatrix}
                \phantom{-}q_1(x) \otimes {\rm id} & \phantom{-}{\rm id} \otimes q_0^*(y) \\
                     \phantom{-}{\rm id} \otimes q_1^*(y) & \phantom{-}q_0(x)\otimes {\rm id}
             \end{smallmatrix}\right)}
           \ar[ddd]_{r_1
              =\left(\begin{smallmatrix} 
                            q_1^{(1)}(x,y) & -q_0^{(1)}(x,y) \\
                            q_1^{(3)}(x,y) & -q_0^{(3)}(x,y) \\[-1ex]
                             \vdots & \vdots
                        \end{smallmatrix} \right)}
        & & & & & 
        {\begin{matrix}Q_0\otimes Q_0^*\\\oplus\\Q_1\otimes Q_1^*\end{matrix}}
        \ar@<0.5ex>[lllll]^{t_0
            =\left(\begin{smallmatrix}
                    \phantom{-}q_0(x)\otimes{\rm id} & -{\rm id}\otimes q_0^*(y) \\
                    -{\rm id}\otimes q_1^*(y) & \phantom{-}q_1^*(x)\otimes {\rm id} 
                 \end{smallmatrix}\right)}
          \ar[ddd]^{r_0
             =\left(\begin{smallmatrix} 
                           {\bf 1} & -{\bf 1} \\
                            q_1^{(2)}(x,y) & -q_0^{(2)}(x,y) \\[-1ex]
                             \vdots & \vdots
                        \end{smallmatrix} \right)}
          \\ \\ \\
        {\Lambda^{\rm odd}V}
          \ar@<0.5ex>[rrrrr]^{
                     \delta+\sigma}
           & & & & & 
           \Lambda^{\rm even}V       
        \ar@<0.5ex>[lllll]^{\delta+\sigma
                   }   
          }
\end{equation}

In order to construct all the entries of the matrices $(r_0,r_1)$ we first observe that the map $\delta$ acts on the module $\Lambda^*V$ as a differential obeying $\delta^2=0$. Thus we can analyse its cohomology. The only non-trivial cohomology is in degree $0$, and it is given by $\CC[x_i,y_i]/(x_1-y_1,\ldots,x_N-y_N)$.

This allows us now to recursively define the entries of the matrices $r_0$ and $r_1$. The first entries, $q_0^{(1)}$ and $q_1^{(1)}$, of the matrix $r_1$ are defined by
\begin{equation} \label{eq:recstart}
     \delta(q^{(1)}_s(x,y)) =  q_s(x) - q_s(y) \ , \qquad s=0,1 \ .
\end{equation}
Note that the right hand sides vanish for $x=y$ and are therefore $\delta$-exact.
Hence these two equations define $q^{(1)}_s$ up to unimportant $\delta$-closed one-forms. 

Then the entries $q^{(k)}_s$ for $k\ge 2$ are given recursively by
\begin{equation} \label{eq:recstep}
     \delta(q^{(k)}_s(x,y)) = 
        -\sum_i A_i(x,y)e_i\wedge q^{(k-2)}_s(x,y)
        +q_s^{(2k-1)}(x,y) q_s(x)-q_s(y)q_{s+1}^{(2k-1)}(x,y) \ ,
\end{equation}
with $q_s^{(0)}=(-1)^{s+1}{\bf 1}$. Similarly to the case $k=1$,
it is straight forward to check that the right hand sides of all the equations \eqref{eq:recstep} are $\delta$-closed. Moreover, due to the fact that there is no non-trivial cohomology at higher degree,
$\delta$-closedness implies $\delta$-exactness. 
Thus the relations~\eqref{eq:recstep} also can be solved recursively to define (again up to unimportant closed contributions) all the other entries of $r_0$ and $r_1$.  

With the definitions of the entries $r_0$ and $r_1$ at hand it is straight forward to check that the diagram \eqref{eq:tachyon} is indeed commutative. In addition, the constant entries in the matrix $r_{0}$ ensure that $(r_1,r_0)$
defines a non-trivial morphism in
 ${\cal H}^0(Q\otimes Q^*, I)$. In fact, it can be identified with the identity morphism in ${\cal H}^0(Q,Q)$.
 
Eventually we need the corresponding morphism $\wt r\in{\cal H}_{\rm orb}^0(\wt Q^\rho \otimes\wt Q^{\rho\, *},\wt I^0)$ in the orbifold category. The orbifold construction for matrix factorisation has been reviewed in Section~\ref{lglgmon}. The upshot is that we obtain the morphism $\wt r$ for the equivariant matrix factorisation $\wt Q$ 
 in the basis which diagonalises the action of the orbifold group $\Gamma\times\Gamma$ simply by 
replacing the variables $x$ and $y$ in $r$ by the matrices $x\,\id$ and $y\,\tau$ introduced in Section~\ref{lglgmon} respectively and then projecting on its $\Gamma\times\Gamma$ invariant part:
\begin{equation}
    \wt{r}(x,y) = \left(r( x\,{\rm id}, y\,\tau )\right)^{\Gamma\times\Gamma}\,.
\end{equation}
%
%
%
%
%
%
\bibliographystyle{mondef}
\bibliography{mondef}
\end{document}